\documentclass[preprint,12pt]{elsarticle} 
\usepackage{amssymb} 
\usepackage{amsmath} 
\usepackage{xcolor}
\usepackage{gensymb}
\usepackage[nolist]{acronym} 
\usepackage[displaymath, mathlines]{lineno} 
\usepackage{booktabs}
\usepackage{physics}%
\usepackage[hyphens]{url}
\pdfoutput=1
\usepackage{graphicx}

\journal{Nuclear Instruments and Methods A} 
\begin{document}

\begin{frontmatter} 
\title{Radio-detection of neutrino-induced air showers:\\ the influence of topography} 
\author[iap]{V. Decoene}
\ead{decoene@iap.fr} 
\author[iap]{N.~Renault-Tinacci}
\author[lpnhe,naoc,iap]{O.~Martineau-Huynh}
\author[subatech]{D. Charrier} 
\author[iap]{K. Kotera} 
\author[naoc]{S.~Le Coz} 
\author[lpc]{V.~Niess} 
\author[iflp,iap]{M.~Tueros}
\author[iap]{A.~Zilles}
 
\address[iap]{Sorbonne Universit\'e, CNRS, UMR 7095, Institut d'Astrophysique de Paris, 98 bis bd Arago, 75014 Paris, France}
\address[lpnhe]{Sorbonne Universit\'e, Universit\'e Paris Diderot, Sorbonne Paris Cit\'e, CNRS/IN2P3, LPNHE, Paris, France} 
\address[naoc]{National Astronomical Observatories of China, Chinese Academy of Science, Beijing 100012, P.R. China} 
\address[subatech]{Universit\'e de Nantes, IMT-Atlantique, CNRS/IN2P3, SUBATECH, Nantes, France} 
\address[lpc]{Universit\'e~Clermont Auvergne, CNRS/IN2P3, LPC, Clermont-Ferrand, France} 
\address[iflp]{Instituto de F\'isica La Plata - CONICET, Argentina}

\begin{abstract} 
Neutrinos of astrophysical origin could be detected through the electromagnetic radiation of the particle showers induced in the atmosphere by their interaction in the Earth.  This applies in particular for tau neutrinos of energies E$>$10$^{16}$\,eV following Earth-skimming trajectories. The $\sim$1$\degree$ beaming of the radio emission in the forward direction however implies that the radio signal  will likely fly above a detector deployed over a flat site and would therefore not be detected. 

We study here how a non-flat detector topography can improve the detection probability of these neutrino-induced air showers. We do this by computing with three distinct tools the neutrino detection efficiency for a radio array deployed over a toy-model mountainous terrain, also taking into account experimental and topographic constraints. We show in particular that ground topographies inclined by few degrees only induce detection efficiencies typically three times larger than those obtained for flat areas for favorable trajectories. We conclude that the topography of the area where the detector is deployed will be a key factor for an experiment like GRAND.
\end{abstract} 

\end{frontmatter} 

\section{Introduction} 
Ultra high energy neutrinos (UHE $\nu$) are valuable messengers of violent phenomena in the Universe (\cite{2016JCAP...12..017F,GRANDWP} and references therein). Their low interaction probability with matter allows them to carry unaltered information from sources located at cosmological distances, but, on the other hand, makes their detection challenging: non-negligible detection probability can be achieved only with large volumes of dense targets. 

At neutrino energies targeted here (E $> 10^{16}$~eV), the Earth is opaque to neutrinos. Therefore only Earth-skimming trajectories yield significant probability of neutrino interaction with matter, leading to a subsequent tau decay in the atmosphere, eventually inducing an extensive air-shower (EAS).
The detection of these EAS has been proposed as a possible technique to search for these cosmic particles\,\cite{Fargion:1999se}. 
The progress achieved by radio-detection of EAS in the last 15 years \cite{Falcke:2005tc,CODALEMA:2009,Tunka:2015,Aab:2015vta,Buitink:2016nkf,Charrier:2018fle} combined with the possibility to deploy these cheap, robust detectors over large areas open the possibility to instrument giant radio arrays designed to hunt for neutrino-induced EAS as proposed by the GRAND project\,\cite{GRANDWP,Ardouin:2011}. 

An EAS emits a radio signal via two well understood mechanisms : the {\it Askaryan effect}~\cite{Askaryan1962,Askaryan1965} and  the {\it geomagnetic effect}~\cite{Kahn1966, Scholten2008}, which add up coherently to form detectable signals in the frequency range between tens to hundreds of MHz. The interplay between these two effects induces an azimuthal asymmetry of the electric field amplitude along the shower axis~\cite{Huege:2016veh,Schroder:2016hrv}.

The nearly perfect transparency of the atmosphere to radio waves, combined with the strong relativistic beaming of the radio emission in the forward direction\,\cite{Alvarez-Muniz:2014dza} make it possible to detect radio signals from air showers at very large distances from their maximum of development $X_{\rm max}$: a $2\times10^{19}$\,eV shower was for example detected by the Auger radio array with a $X_{\rm max}$ position reconstructed beyond $100$\,km from shower core~\cite{Aab:2018ytv}. This is obviously an important asset in favor of radio-detection of neutrino-induced air showers.  

The strong beaming of the radio signal also implies that the topography of the ground surface may play a key role in the detection probability of the induced EAS.
The primary objective of this article is to perform a quantitative study of the effects of ground elevation on the detection probability of neutrino-induced air showers. To do this, we use a toy configuration where a radio array is deployed over a simplified, generic topography. We compute the response of this setup to neutrino-induced showers with three different simulation chains, ranging from a fast and simple estimation using a parametrization of the expected signal amplitude, to a detailed and time consuming Monte-Carlo. This is motivated by the fact that full Monte-Carlo tools are CPU-intensive treatments, to the point of becoming prohibitive when it comes to simulating radio detection by large antenna arrays. The secondary purpose of this paper is therefore to determine if reliable results can be obtained with faster treatments than full Monte-Carlo simulation codes. 

In section \ref{principle} we present the general principle of our study, in section \ref{sim} we detail the implementation of the three simulation chains used, and finally in section \ref{res} we discuss the results.

\section{General principle} 
\label{principle}

\begin{figure*}[t]
    \begin{center}
    \centering
    \includegraphics[width=\textwidth,trim=0cm 9cm 0cm 0cm,clip=true]{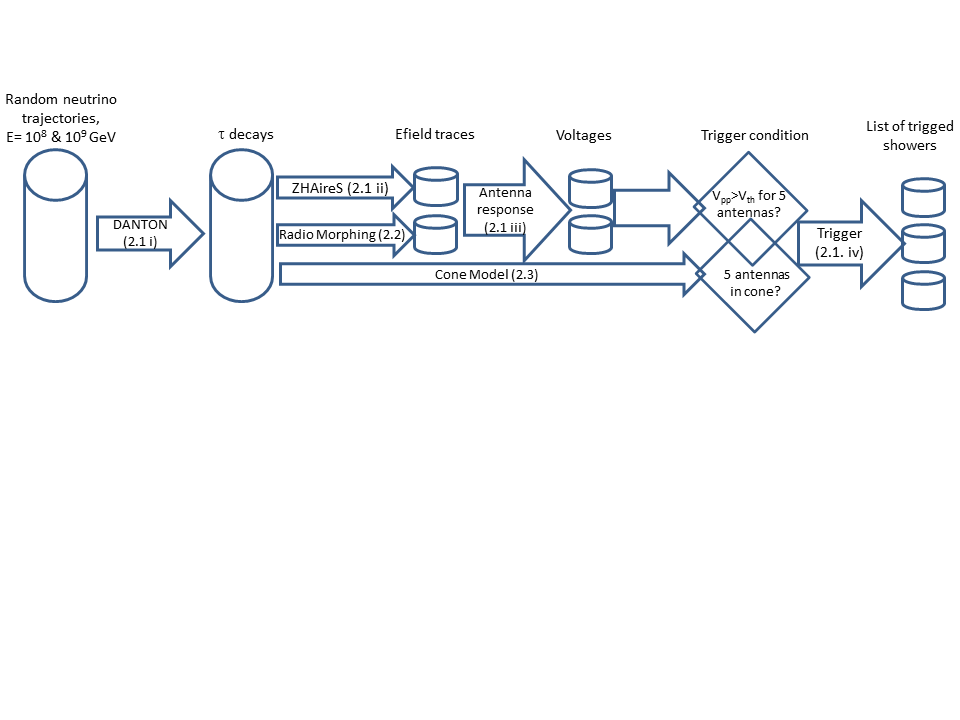}
    \caption{General structure of the three simulation chains ({\it microscopic, Radio Morphing and Cone} models) used in this study. The sections where their various elements are described are indicated in parenthesis. The trigger condition for the {\it microscopic} and {\it Radio Morphing} methods is fulfilled if five antennas or more with peak-peak voltages larger than a threshold value $V_{th}$ set to 5 times (conservative) or twice the minimal background noise level. For the {\it Cone model}, the trigger condition is fulfilled if five antennas or more are within the volume of the cone modeling the shower radio emission. }
    \label{principle_schema}
    \end{center}
\end{figure*}

Three simulation chains are used in this study. Their general principles are presented in sections \ref{end2end} to \ref{coneprinciple}, and summarized in Figure\,\ref{principle_schema}. In section \ref{toymodeldescription}, we present the toy detector configuration used for the study.

\subsection{End-to-end microscopic simulation}
\label{end2end}

The first simulation chain consists of four independent steps:
\begin{enumerate}[(i)]   
\item We produce a fixed number of tau decays induced by cosmic tau-neutrinos ($\nu_{\tau}$) interacting in a spherical Earth. This is done for two neutrino energies ($E_{\nu} = 10^9$ and $10^{10}$\,GeV) with a dedicated Monte-Carlo engine:  DANTON\,\cite{DANTON:note,DANTON:GitHub}, further described in section \ref{danton}.
\item We compute the electromagnetic field induced at the location of the detection units by the showers generated by these tau decays. This is done through a full {\it microscopic simulation} of the particles in the EAS and of the associated electromagnetic radiation using the ZHAireS\,\cite{ZHAireS} simulation code  (see section \ref{zhaires} for details).
\item The voltage induced by the radio wave at the antenna output is then computed using a modelling of the GRAND {\sc HorizonAntenna}\,\cite{GRANDWP} performed with the NEC4\,\cite{NEC4} code. This is detailed in section \ref{horant}.

\item 
If the peak-to-peak amplitude of the output voltage exceeds the defined threshold for five antennas or more, then the neutrino is considered as detected (see section \ref{trig} for more details). This threshold value is either twice (aggressive scenario) or five times the minimal noise level (conservative scenario).
\end{enumerate}

\subsection{Radio-Morphing}
\label{radiomorphing}
Monte-Carlo simulations of the electric field provide the most reliable estimate of the detection probability of a shower, and are therefore used as a benchmark in this work. They however require significant computational resources: the CPU time is mainly proportional to the number of simulated antennas and can last with ZHAireS up to $\approx72$\,h on one core for $1000$ antennas given our simulation parameters.  

An alternative simulation chain therefore uses the so-called {\it Radio Morphing} method \cite{Zilles:2018kwq} instead of ZHAireS for the electric field computation. {\it Radio Morphing} performs a very fast, semi-analytical computation of the electric field (see section \ref{rm} for details). The antenna response and the trigger computation are simulated in the same way as for the {\it microscopic simulation} chain. The gain in computation times allows to study a larger number of configurations than with the {\it microscopic} approach.  

\subsection{Cone Model}
\label{coneprinciple}
Even if significantly faster than the microscopic method, the {\it Radio Morphing} treatment still requires that the antenna response is computed, and thus implies that hundreds of time traces for electric field and voltage need to be handled for each simulated event. A third, much lighter method is therefore used in this study. It is based on a geometric modeling of the volume inside which the electromagnetic field amplitude is large enough to trigger an antenna. We give to this volume the shape of a cone, oriented along the shower axis, with its apex placed at the $X_{\rm max}$ position, half-angle $\Omega$ and height $H$. Values of $\Omega$ and $H$ depend on shower energy, and are adjusted from ZHAireS simulations (see section \ref{cone} for details). A shower is considered as detected if at least five antennas are within the cone volume. A similar {\it Cone Model} was used to compute the initial neutrino sensitivity of the GRAND detector\,\cite{Martineau:2016yj}.Being purely analytical, this method produces results nearly instantaneously and requires only minimal disk space and no specific simulation software, an attractive feature when it comes to perform simulation for thousands of detection units covering vast detection areas. \\

\subsection{Toy detector configuration}
\label{toymodeldescription}

\begin{figure*}[t]
    \centering
    \includegraphics[width=0.8\textwidth]{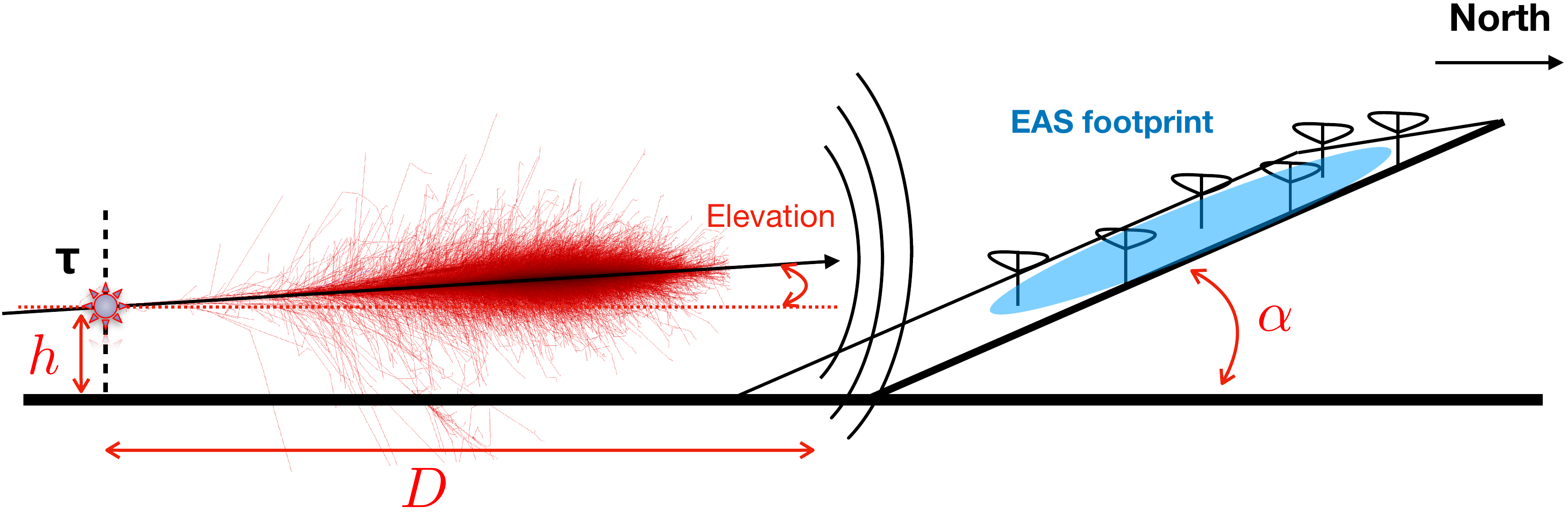}
    \caption{Layout of the toy-setup considered in this study. A tau particle decays at a location represented as a star, producing an air shower. The radio signal emitted by the shower impinges the detector plane, tilted by an angle $\alpha$ from the horizontal. The intersection between the detector plane and the horizontal plane is set at a horizontal distance $D$ from the decay point. The parameter $D$ is therefore a measurement of the amount of free space in front of the detector.}
    \label{cone_schema}
\end{figure*}

The detector considered in this study is presented in Figure~\ref{cone_schema}. It is a rectangular grid with a step size of $1000$\,m between neighbouring antennas. This large step size is a distinct feature of the envisioned dedicated radio array for the detection of neutrino-induced air showers\,\cite{Alvarez-Muniz:2014dza}. It is a compromise between the need for very large detection areas imposed by the very low event rates expected for one part, and the instrumental and financial constraints which limit the number of detection units for the other.

In our study we use a simplified, toy setup configuration where the antenna array is deployed over a plane inclined by an adjustable angle $\alpha$ (also called "slope" in the following) with respect to an horizontal plane. We restrict our treatment to showers propagating to the North, i.e. directly towards the detector plane. For other directions of propagation, the size of the shower footprint on ground ---hence its detection probability---  would directly depend on the width of the detector plane. Defining a specific value for this parameter would be highly arbitrary, given the great diversity of topographies existing in reality. For showers propagating towards the detector however, the shower footprint is aligned with the detector longitudinal axis (see Figure \ref{cone_schema}), and the detector width then has a negligible effect on the shower detection efficiency.  This motivates our choice to restrict ourselves to this single direction of propagation. The horizontal distance between the tau decay point and the  foot of the detector, $D$, can be understood as the amount of empty space in front of the detector over which the shower can develop and the radio signal propagate. It is therefore closely related to the topography of the detection site. The reference ground elevation is chosen to be 1500~m above sea level (a.s.l.). A maximum altitude of $4500$\,m a.s.l. is set for the antennas, as larger elevation differences are unrealistic. The vertical deviation due to Earth curvature can be estimated by $2\delta h \approx R_{\rm earth} (L/R_{\rm earth})^2$\,km, where $L << R_{\rm earth}$ is the longitudinal distance between the maximum development of the shower and the observer. For $L=50$\,km, we find $\delta h < 100$\,m. A flat Earth surface is therefore assumed in this toy setup configuration.

The slope $\alpha$  and the distance $D$ are the two adjustable parameters of the study.  Values of $\alpha$ vary from 0 to 90$\degree$ and $D$ ranges between 20 and 100\,km, covering a wide variety of configurations. As will be detailed in section \ref{toymodelres},  larger values of $D$ are irrelevant because most showers would then fly over the detector (see Figure\,\ref{fly_above} in particular), an effect that would furthermore be amplified if the Earth curvature was taken into account. Values of $\alpha$ larger than 30$\degree$ are also not realistic, because steeper slopes are not suitable to host a detector, but they are included in our study for the sake of completeness, and because these extreme cases will help us interpret the results of the study.

For each  pair of values $(\alpha, D)$, we process the two sets of tau decays of energies $E_{\nu}=10^9$\,GeV and $E_\nu=10^{10}$\,GeV with the three methods {\it microscopic}, {\it Radio Morphing} and {\it Cone Model}. We then use the fraction of tau decays inducing a trigger by the detector to perform a relative comparison between $(\alpha, D)$ configurations. This treatment allows to directly assess the effect of topography on neutrino-induced shower detection efficiency ---the purpose of this paper--- for a reasonable amount of computing time, given the large number of configurations ($\alpha$, D) considered in this study. \\

\section{Computational methods} 
\label{sim}

We present in the following the implementation of the methods described in sections \ref{end2end} to \ref{coneprinciple}.

\subsection{Production of the shower progenitors}
\label{danton}

\begin{figure*}[tb]
    \centering
    \includegraphics[width=0.32\textwidth]{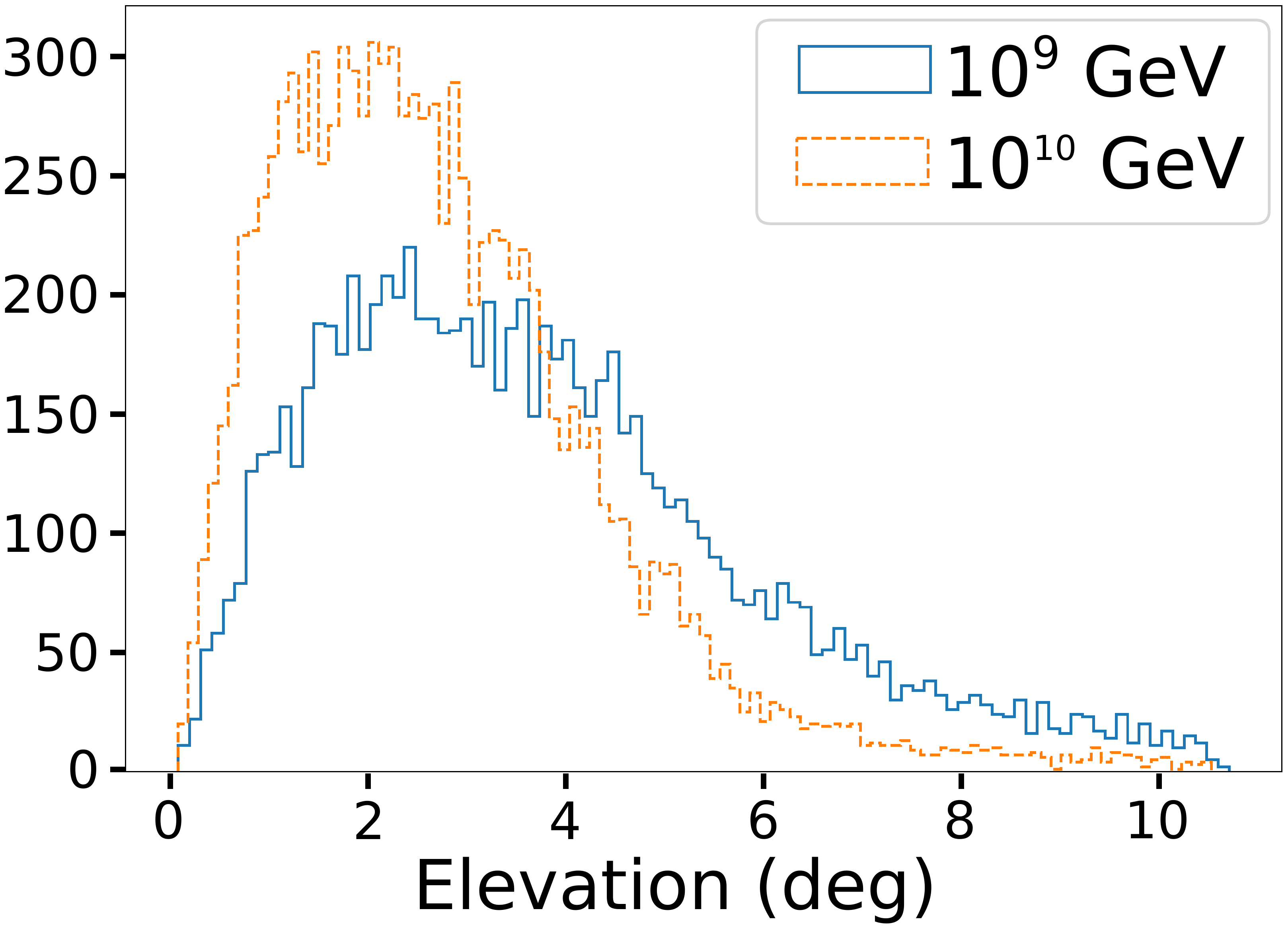}
    \includegraphics[width=0.32\textwidth]{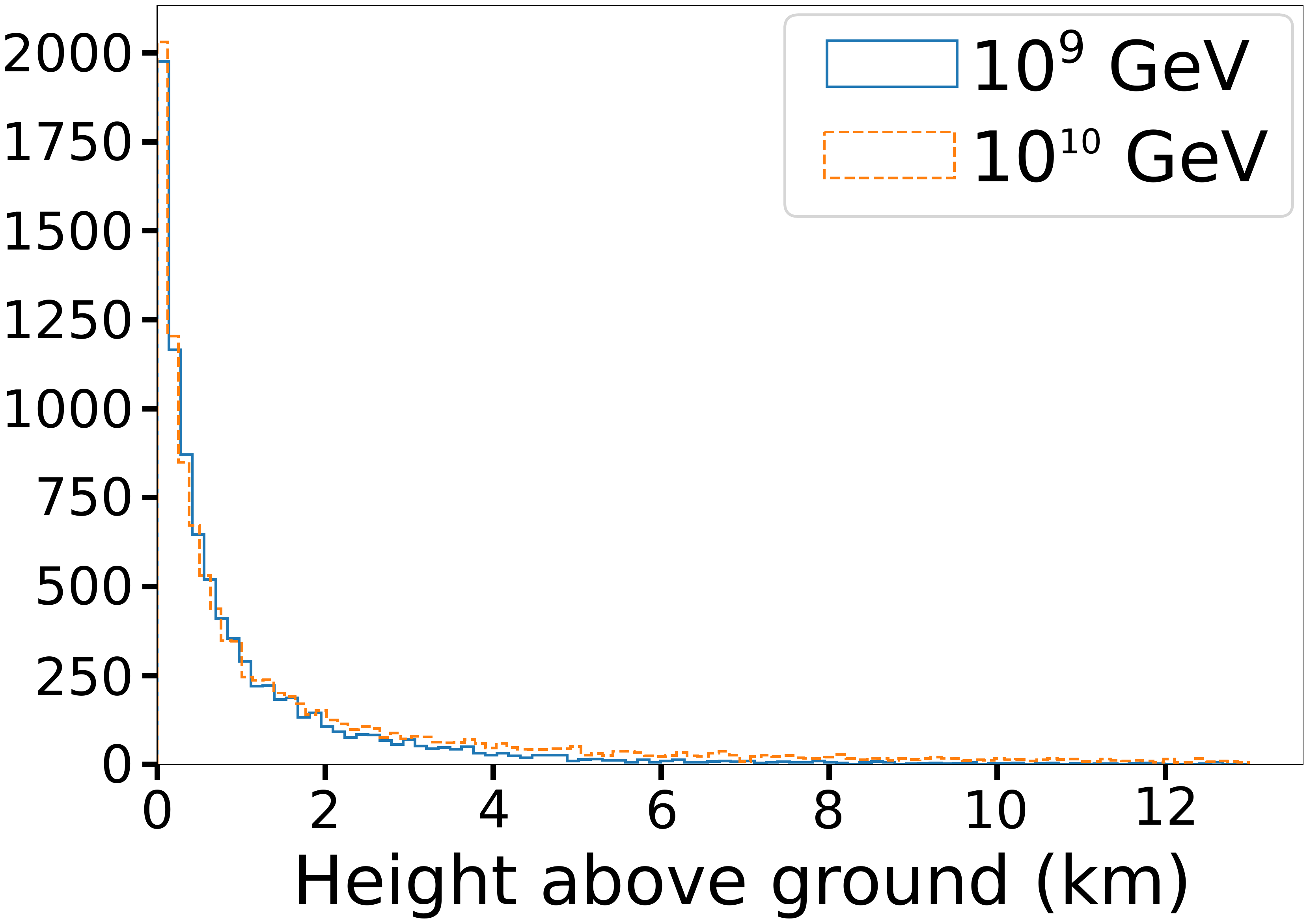}
    \includegraphics[width=0.32\textwidth]{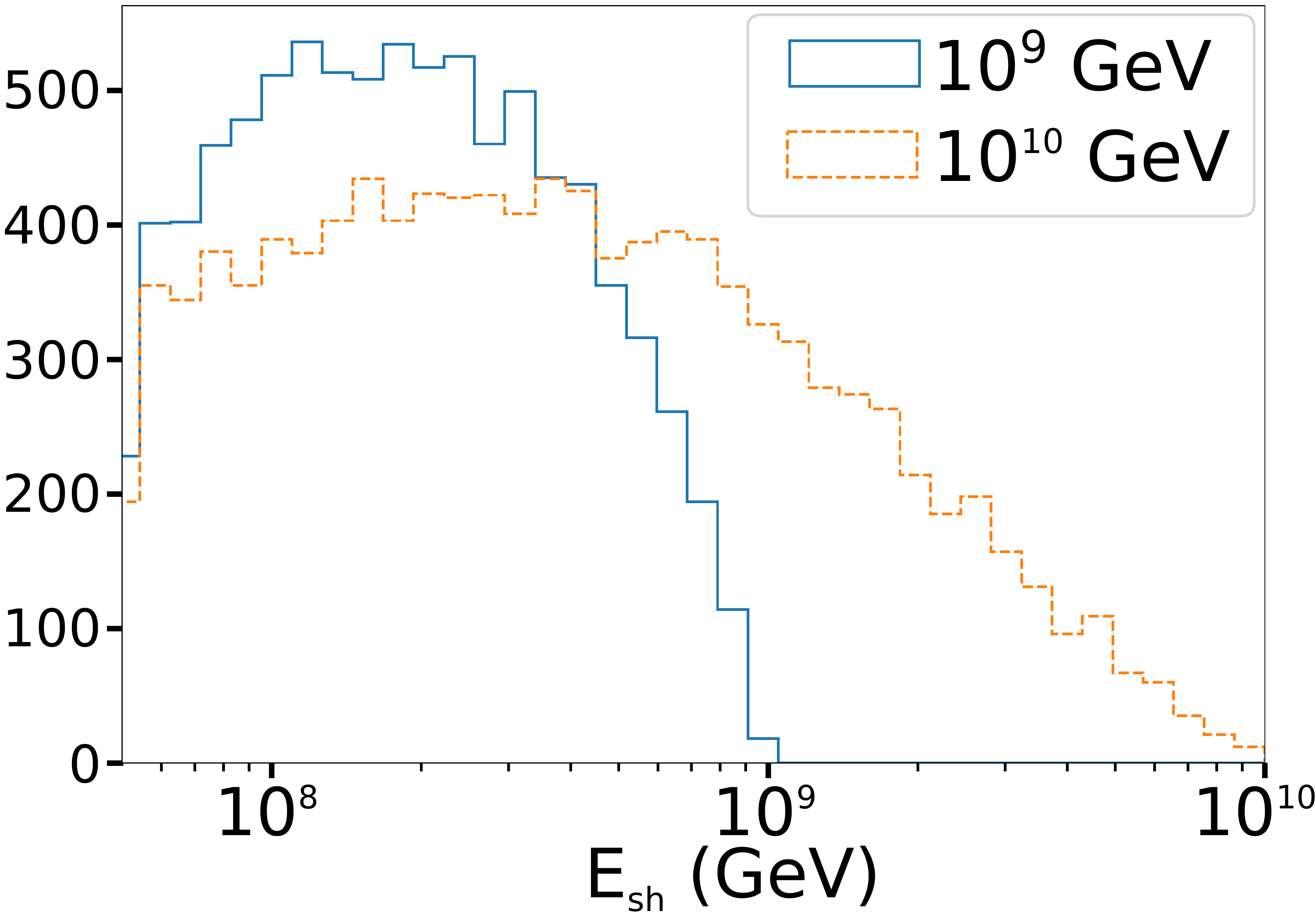}
    \caption{Distributions of tau decay elevation angles of particle trajectory measured with respect to a horizontal plane ({\it left}), height above ground at tau decay point ({\it center}) and shower energy ({\it right}) for the two sets of primary $\nu_{\tau}$ energy considered in this study.}
    \label{primary_distrib}
\end{figure*}

The production of the shower progenitors was performed with the DANTON software package \cite{DANTON:note, DANTON:GitHub}. DANTON simulates interactions of tau neutrinos and tau energy losses. It produces results compatible with similar codes\,\cite{2018PhRvD..97b3021A}. Additionally DANTON offers the possibility to run simulations in backward mode (i.e. from tau decay upwards, with appropriate event weight computations), an attractive feature for massive simulations, and it also allows us to take into account the exact topography of the Earth surface\,\cite{2019arXiv190403435N}. It is however operated here in forward mode, i.e. as a classical Monte-Carlo. The primary neutrino source is set as mono-energetic and isotropic. A spherical Earth is used with a density profile given by the Preliminary Reference Earth Model (PREM)\,\cite{PREM}, but with the sea layer replaced by Standard Rock\,\cite{StandardRock}. Two energy values are used for the primary neutrino flux: $E_{\nu}=10^{9}$~GeV and $10^{10}$~GeV. The characteristics of the tau lepton resulting from the interaction of the neutrino with the Earth and of all the particles produced during the decay of the tau in the atmosphere are also computed: decay position, list of products and their associated momenta.

For this study one million primary neutrinos were simulated per energy value. Those inducing tau decays in the atmosphere were then selected if the subsequent showers had energies above $5 \times 10^{7}~{\rm GeV}$, because lower energies can hardly lead to detection for such a sparse array\,\cite{CODALEMA:2009, Huege:2016veh}. In Figure \ref{primary_distrib}, we show the distribution in energy, elevation angle and height of the two sets of tau decays. Among the surviving set, $100$ were randomly chosen for each energy. This value is a good compromise between computation time and statistical relevance.
%

\subsection{Simulation of the electric field}
\subsubsection{Microscopic method}
\label{zhaires}
In the {\it microscopic} method, the extensive air showers initiated by the by-products of the tau decay, and the impulsive electric field induced at the antenna locations were simulated using the ZHAireS software \cite{ZHAireS}, an implementation of the ZHS formalism \cite{ZHS} in the AIRES \cite{AIRES} cascade simulation software. To allow for geometries where cascades are up-going and initiated by multiple decay products, we implemented a dedicated module called RASPASS (Radio Aires Special Primary for Atmospheric Skimming Showers) in the ZHAireS software.

\subsubsection{Radio Morphing}
\label{rm}

{\it Radio Morphing}~\cite{Zilles:2018kwq} is a semi-analytical method for a fast computation of the expected radio signal emitted by an air shower. The method consists in computing the radio signal of any {\it target} air shower at any target position by simple mathematical operations applied to a single {\it generic} reference shower. The principle is the following: 
\begin{enumerate}[i)]
    \item The electromagnetic radiation associated with the {\it generic} shower is simulated using standard microscopic tools at positions forming a 3D mesh.
    \item For each {\it target} shower, the simulated signals are scaled by numerical factors which values depend analytically on the energy and geometry of the {\it target} and {\it generic} showers.
    \item The {\it generic} 3D mesh is oriented along the  direction of propagation of the target shower.
    \item The electromagnetic radiation expected at a given {\it target} position is computed by interpolation of the signals from the neighbouring positions of this 3-D mesh.
\end{enumerate}
 This technique lowers the required CPU time of at least two orders of magnitude compared to a standard simulation tool like ZHAireS, while reproducing its results within $\sim 25\%$ error in amplitude\,\cite{Zilles:2018kwq}.

\subsection{Antenna response}
\label{horant}
In order to compute the voltage generated at the antenna output for both {\it microscopic} and {\it Radio Morphing} methods, we choose in this study the prototype antenna for the GRAND project: the {\sc HorizonAntenna}\,\cite{GRANDWP}.
It is a bow-tie antenna inspired from the {\it butterfly antenna}\,\cite{Charrier:1999ARENA} developed for the CODALEMA experiment\,\cite{Escudie:2019ni}, later used in AERA\,\cite{Abreu:2012pi} and adapted to GRANDProto35\,\cite{GP35:2017}. As for GRANDProto35, three arms are deployed along the East-West, South-North and vertical axes, but the radiating element is half its size to better match the $50-200$\,MHz frequency range considered for GRAND. As the {\it butterfly antenna}, the {\sc HorizonAntenna} is an active detector, but in the present study, we simply consider that the radiator is loaded with a resistor $R = 300\,\Omega$, with a capacitor $C = 6.5 \times 10^{-12}\,$F and inductance $L = 1\,\mu$F in parallel. The {\sc HorizonAntenna} is set at an height of 4.5\,m above ground in order to minimize the ground attenuation of the radio signal.

The equivalent length ${\vec{l}_{eq}}^k$ of one antenna arm $k$ (where $k$ = EW, NS, Vert) is derived from NEC4\,\cite{NEC4} simulations as a function of wave incoming direction ($\theta$, $\phi$) and frequency $\nu$. The voltage at the output of the resistor $R$ loading the antenna arm is then computed as:
\begin{equation}
\label{vant}
V^k(t) = \int {\vec{l}_{eq}}^k(\theta, \phi, \nu) \cdot \vec{E}(\nu) e^{2i \pi \nu t} d\nu
\end{equation}
where $\vec{E}(\nu)$ is the Fourier transform of the radio transient $\vec{E}(t)$ emitted by the shower.

The equivalent length was computed for a vertical antenna deployed over a flat, infinite ground. The ground slope of the toy setup can then be accounted for by a simple rotation of this system by an angle $\alpha$, which translates into a wave effective zenith angle $\theta^*=\theta-\alpha$, to be used in Eq. \ref{vant}. 

\subsection{Trigger}
\label{trig}
The last step of the treatment consists in determining whether the shower could be detected by the radio array. For this purpose, we first apply a Butterworth filtering of order $5$ to the voltage signal in the $50-200$\,MHz frequency range. This mimics the analog system that would be applied in an actual setup in order to filter out background emissions outside the designed frequency range. 

Then the peak-to-peak amplitude of the voltage $V_{\rm pp}$ is compared to the level of stationary background noise $\sigma_{\rm noise}=15\mu V$, computed as the sum of Galactic and ground contributions (see \cite{GRANDWP} and \cite{Charrier:2018fle} for details). If $V_{\rm pp}\geq N \sigma_{\rm \rm noise}$, then we considered that the antenna has triggered. Here $N$ = 2 in an aggressive scenario, which could be achieved if innovative triggering methods\,\cite{FuhrerARENA:2018, Erdmann:2019nie} were implemented, and $N$ = 5 in a conservative one.

If at least five antennas trigger on a same shower, then we consider it as detected.

\subsection{Cone Model}
\label{cone}
The {\it Cone Model} proposes to describe the volume inside which the electromagnetic radiation is strong enough to be detected as a cone, characterized by a height and an opening angle varying with shower energy. The {\it Cone Model} allows for a purely analytical computation of the radio footprint at ground, and thus provides a very fast evaluation of the trigger condition, while it also allows for an easier understanding of the effect of ground topography on shower detection.

The parametrization of the cone height and opening angle as a function of shower energy needs to be computed once only for a given frequency range. This was done as follows for the $50-200$\,MHz band considered in this study: 

\begin{figure}[tb]
    \centering
    \includegraphics[width=0.9\columnwidth]{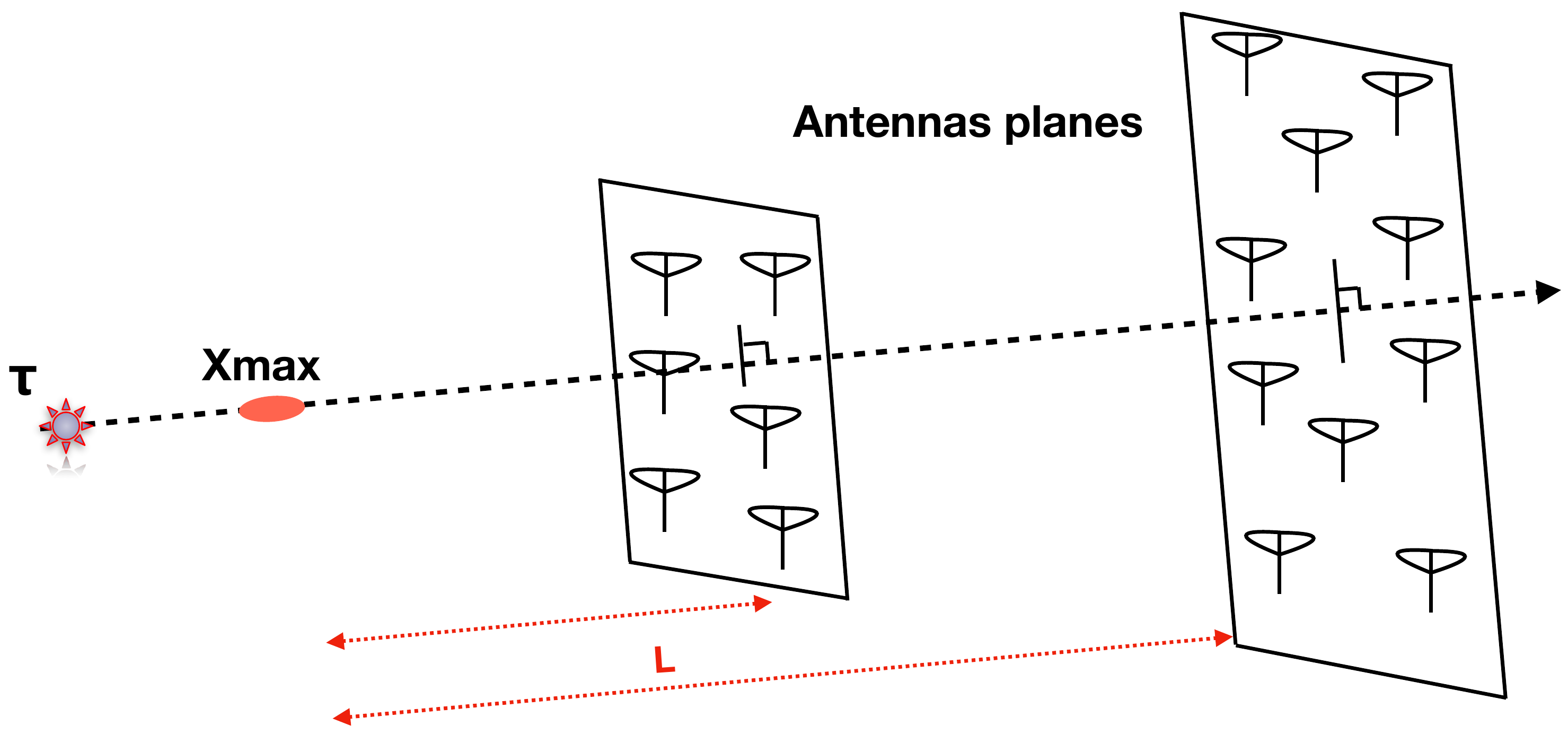}
    \caption{Position of the planes used to parametrize the {\it Cone Model}. These are placed perpendicular to the shower axis, at various longitudinal distances $L$ from $X_{\rm max}$. See section \ref{cone} for details.}
    \label{config}
\end{figure}

\begin{enumerate}
\item We simulate with the ZHAireS code the electric field from one shower at different locations set at fixed longitudinal distances $L$ from the $X_{\rm max}$ position (see Figure \ref{config} for an illustration). Values $L>$100\,km are not simulated because the maximal value $D$ = 100\,km chosen in our study for the distance between the tau decay point and the basis of the detector (see section \ref{toymodeldescription}) makes it unnecessary. As the $X_{\rm max}$ position is reached $\sim$15\,km after the decay, a distance L = 100\,km allows to simulate radio signals over a detector depth of 15\,km at least. This is, in the majority of cases, enough to determine if the shower would be detected or not. 
\item In each of these antenna planes, identified by an index $j$ in the following, we compute the angular distance between the antennas and the shower core. We determine the maximal angular distance to the shower core $\Omega^j$ beyond which the electric field drops below the detection threshold, set to 2 (aggressive) or 5 (conservative) times the value of $E_{\rm rms}$, the average level of electromagnetic radiation induced by the Galaxy is computed as:
\begin{equation}
    {E_{\rm rms}}^2 = \frac{Z_0}{2}\int_{\nu_0}^{\nu_1}\int_{2\pi}B_{\nu}(\theta,\phi,\nu)\sin(\theta) d\theta d\phi d\nu
\end{equation} 
where $B_{\nu}$ is the spectral radiance of the sky, computed with GSM \cite{Gal:2016} or equivalent codes, $Z_0=376.7$\,$\Omega$ the impedance of free space, and [$\nu_0,\nu_1$] the frequency range considered for detection. Here we choose $\nu_0=50$\,MHz and $\nu_1$=200\,MHz, the frequency range of the {\sc HorizonAntenna}. 
The factor $1/2$ arises from the projection of the (unpolarized) Galactic radiation along the antenna axis. We find $E_{\rm rms}$ = 22\,$\mu$V/m. Defining a detection threshold on the electric field amplitude as done here ---rather than the voltage at antenna output as usual--- allows to derive results that do not depend on a specific antenna design. It is however not precise: by construction, the details of a specific antenna response and its dependency on the direction of origin of the signal are neglected here, and only the average effect is considered. The {\it Cone Model} is therefore only an approximate method.

The distribution of the electric field amplitudes as a function of the angular distance to the shower axis is shown for illustration in Figure \ref{omega_trig} for the plane $j$ located at a longitudinal distance $L=59$\,km. As the Cherenkov ring induces an enhancement in the amplitude profile for $\Omega\sim1$\degree, we actually compute two values of the angle $\Omega^j$:  $\Omega^j_{\rm min}$ and $\Omega^j_{\rm max}$, thus defining the angular range inside which the electric field amplitude is above the detection threshold.

\begin{figure}[tb]
    \centering
    \includegraphics[width=0.9\columnwidth]{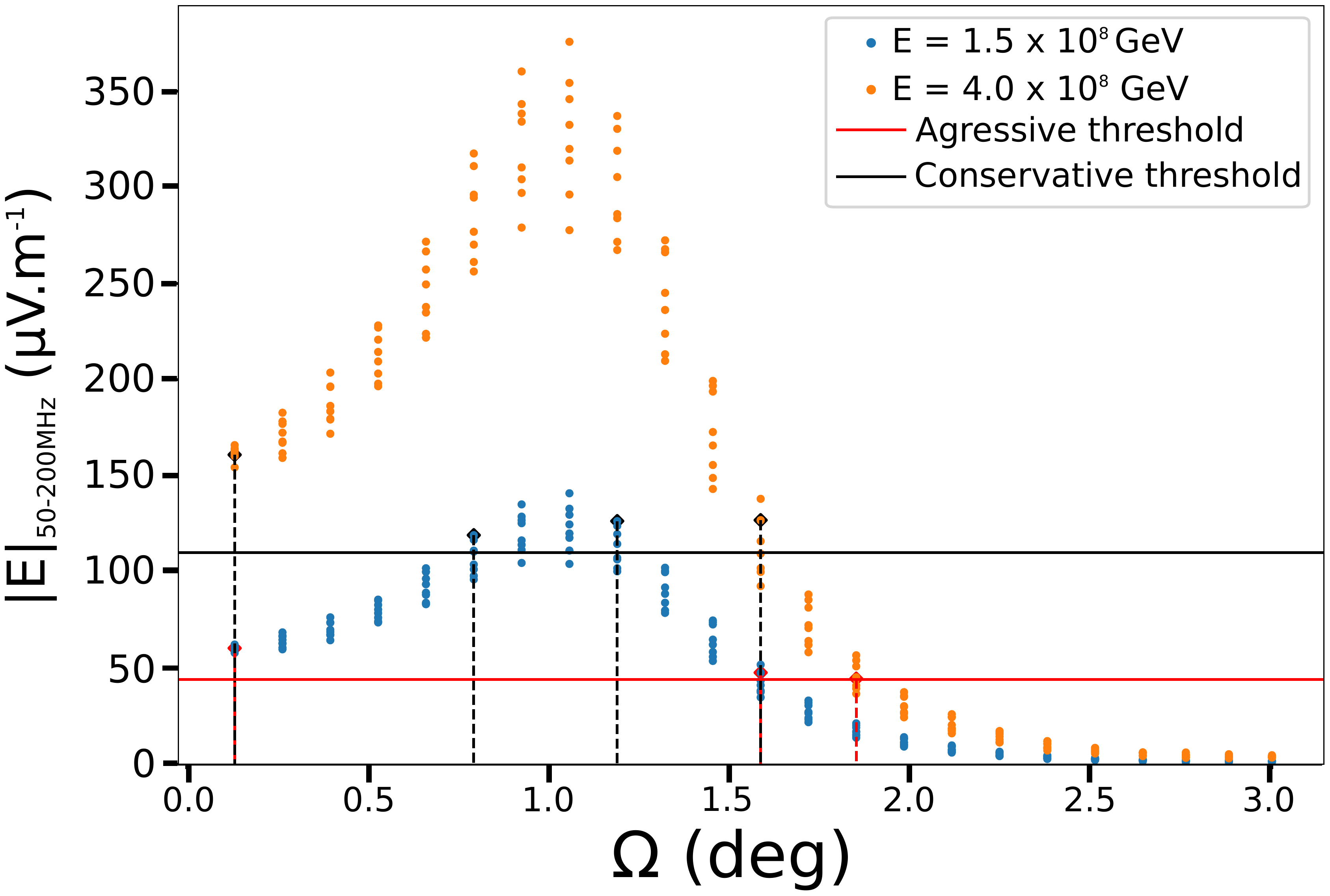}
    \caption{Distribution of the electric field amplitude produced with ZHAireS} as a function of $\Omega$, the angular distance to the shower axis, for antennas located at a longitudinal distance of $59$\,km from $X_{\rm max}$. The amplitude dispersion at a given $\Omega$ value is due to interplay between Askaryan and geomagnetic effects leading to an azimuthal asymmetry of the signal amplitude. For a shower energy $E=1.5\times10^8$\,GeV (in blue) from the $E_{\nu}=10^9$\,GeV dataset, we find for instance ($\Omega^j_{\rm min}$; $\Omega^j_{\rm max}$) =  (0.1$\degree$; 1.7$\degree$) in the aggressive case and (0.8$\degree$; 1.2$\degree$) in the conservative one. 
    \label{omega_trig}
\end{figure}

\item The value of $\Omega^j$ does not vary significantly with $L$ (see Figure \ref{omegaVSdist}). This validates the choice of a conical model for the trigger volume and allows to derive a single set of values ($\Omega_{\rm min};\Omega_{\rm max}$)=($\langle \Omega^j_{\rm min} \rangle$; $\langle \Omega^j_{\rm max} \rangle$) for one specific energy. 
 
\begin{figure}[tb]
    \centering
    \includegraphics[width=0.9\columnwidth]{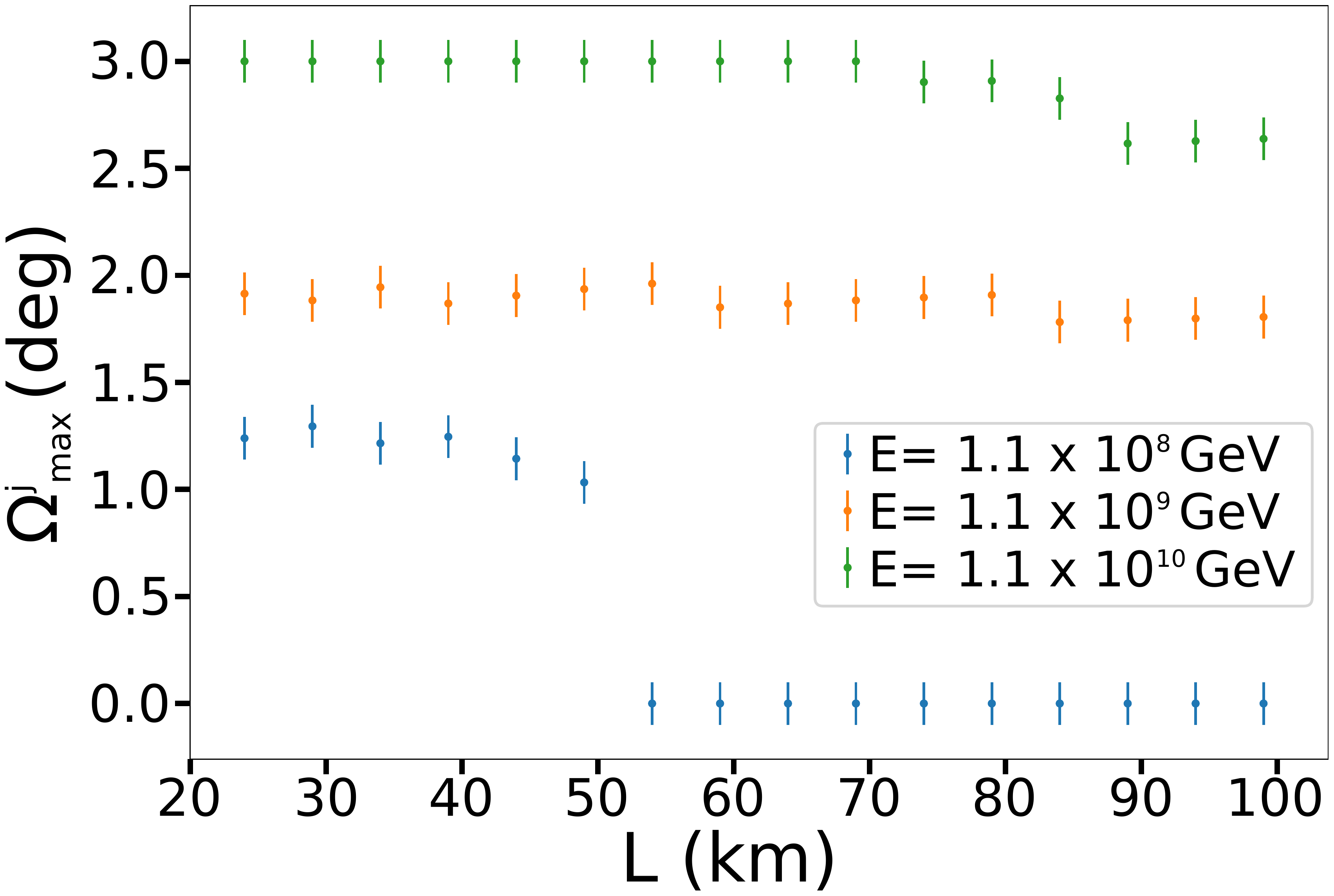}
    \caption{Angular distances $\Omega^j_{\rm max}$ computed following the method presented in Figure \ref{omega_trig} as a function of longitudinal distance for various shower energies. $\Omega_{\rm max}$ measures the maximum opening angle of the cone describing the triggering volume, while index $j$ identifies the simulation plane perpendicular to the shower axis (see Figure \ref{config}).  Here only the conservative case in shown.} The angle value varies marginally over the full  range of longitudinal values considered for shower energies $E = 1.1\times10^9$ and $1.1\times10^{10}$\,GeV, validating the choice of a cone model ---with fixed opening angle $\Omega_{\rm max}=<\Omega_{\rm max}^j>$--- for the trigger volume modeling. For $E = 1.1\times10^8$\,GeV, $\Omega$ drops to 0 for $L>50$\,km because the cone height $H$ is equal to 50\,km in the conservative case (see Figure \ref{dist_E}). Similar treatment is applied to determine $\Omega_{\rm min}$. 
    \label{omegaVSdist}
\end{figure}

\item A similar procedure is applied to determine the cone height $H$, set to be equal to the longitudinal distance $L$ up to which the signal is strong enough to be detected. 

\begin{figure}[tb]
    \centering
    \includegraphics[width=0.9\columnwidth]{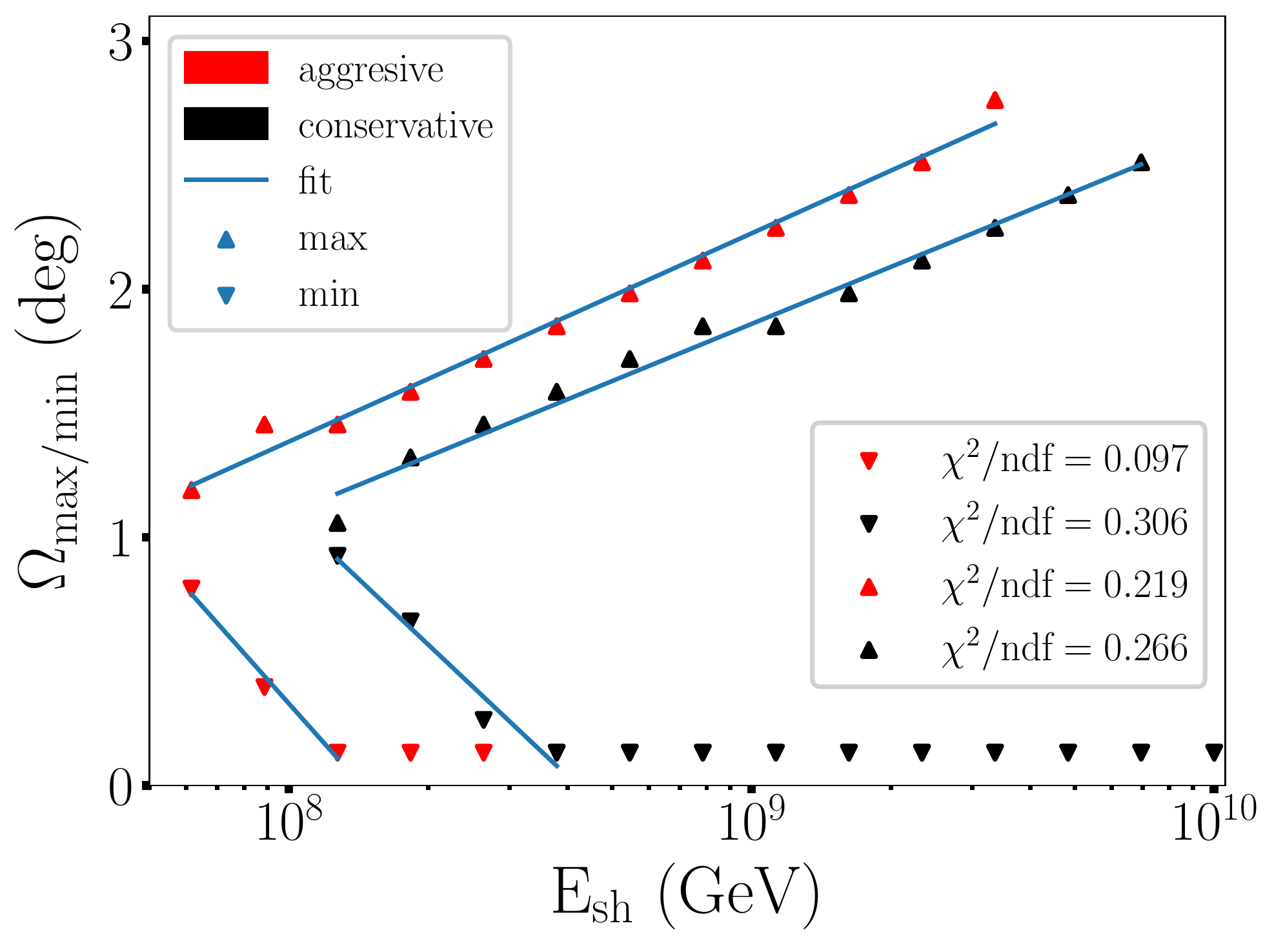}
    \caption{Angles $\Omega_{\rm max}$ and $\Omega_{\rm min}$  as a function of shower energy $E_{\rm sh}$; and fit by Eq. \ref{eq:fitOmega}. Angles $\Omega_{\rm max}$ and $\Omega_{\rm min}$ define the inner and outer boundaries of the hollow cone and are obtained by averaging the values $\Omega_{\rm max}^j$ and $\Omega_{\rm min}^j$ (see Figure \ref{omegaVSdist})}. At the highest energies, $\Omega_{\rm min}$ drops down to $0$\degree, implying that the radio signal is above the detection threshold for all angular distances $\Omega \leq \Omega_{\rm max}$.
    \label{omega_E}
\end{figure}

\begin{figure}[tb]
    \centering
    \includegraphics[width=0.9\columnwidth]{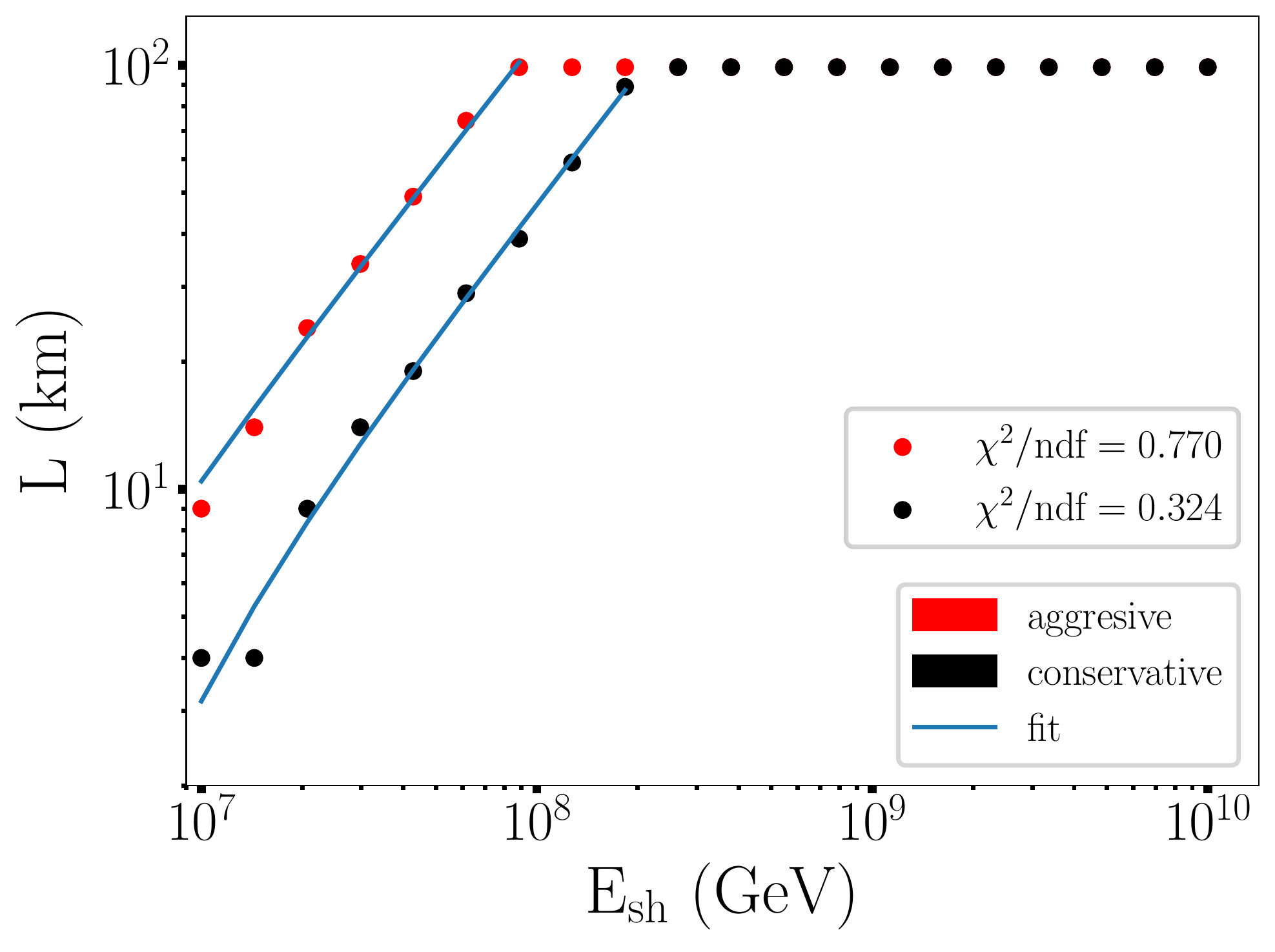}
    \caption{Cone height $H$ as a function of shower energy $E_{\rm sh}$ and fit by Eq. \ref{eq:fitL}. Cone height saturates at values $H$\,=\,100\,km, because the antenna planes used to parametrize the {\it Cone Model} do not extend beyond this value. However points with values $H<100$\,km suffice to demonstrate that the cone heights scale linearly with energy, as one would naturally expect, since the electric field amplitude also scales linearly with energy. Cone height values $H>100$\,km are therefore extrapolated from the fit given in Eq. \ref{eq:fitL}.}
    \label{dist_E}
\end{figure}

\begin{table}[ht]
\begin{center}
\caption{Parameters for the fitting functions given in Equations~\ref{eq:fitL} and~\ref{eq:fitOmega}, for aggressive and conservative thresholds and maximal and minimum $\Omega$ angles. Parameters $a$ and $b$ are in km, $c$ and $d$ in degrees. }
\label{tab:fitting_values}
\hspace{0.4cm}
\begin{tabular}{ccccccc}
\toprule threshold &$a$ & $b$ & $\Omega$ & c & d\\ 
\midrule
\midrule aggressive& 109 $\pm$ 15 & 116 $\pm$ 3 & min & 0.20 $\pm$ 0.02 &-2.4 $\pm$ 0.2\\
&&& max & 1.3 $\pm$ 0.2 &1.00 $\pm$ 0.02 & \\
\midrule conservative & 42 $\pm$ 7 & 48 $\pm$ 1 & min & 1.2 $\pm$ 0.2 & -2.2 $\pm$ 0.2 \\
&&& max & 1.0 $\pm$ 0.3 &0.80 $\pm$ 0.03 \\
\midrule
\bottomrule
\end{tabular}
\end{center}
\end{table}

\item We repeat the treatment for various shower energies $E_{\rm sh}$ by rescaling the signals amplitudes and thus obtain the distributions $\Omega(E_{\rm sh})$ and  $H(E_{\rm sh})$ shown in Figures \ref{omega_E} and \ref{dist_E}. We fit these distributions for shower energies larger than $3 \cdot 10^{7}$\,GeV with analytic functions given by
\begin{eqnarray}
    H|_{\rm 50-200MHz} =& a\, +\ b\, \left(\frac{E_{\rm sh}-10^{17} {\rm eV}}{\rm 10^{17} eV}\right), 
    \label{eq:fitL} \\
    \Omega|_{\rm 50-200MHz} =& c\, +\ d\, \log{\left(\frac{E_{\rm sh}}{\rm 10^{17}eV}\right)}. \label{eq:fitOmega} 
\end{eqnarray}
with $E_{\rm sh}$ expressed in eV in the formulas. Numerical values of $a,b,c,d$ are given in Table~\ref{tab:fitting_values}.

The three parameters $\Omega_{\rm min}$, $\Omega_{\rm max}$ and $H$ allow to define a hollow cone, with an apex set at the shower $X_{\rm max}$ location and oriented along the shower axis. Any antenna located inside this volume is supposed to trigger on the shower according to the {\it Cone Model}.

As mentioned in the introduction, the interplay between the geomagnetic effect and the charge excess induces an asymmetry on the electric field amplitude as a function of antenna angular position w.r.t. the shower core. This can be seen on Figure \ref{omega_trig}, for instance, where the dispersion in field strength at a given angular distance is the exact illustration of this phenomenon. The {\it Cone Model} however assumes a rotation symmetry around the shower axis and thus neglects this asymmetry. This is still acceptable if we are only interested in the average number of triggered antennas by the shower ---which is the case here--- and not in the amplitude pattern of the radio signal. 
\end{enumerate}

Once this parametrization is completed, the {\it Cone Model} is applied to the selected set of tau decays: the values of the cone parameters are computed for the energy and geometry of each shower and the intersection between the resulting cone volume and the detection area is calculated. If at least five antennas fall within this intersection, then we consider that the shower is detected.

\section{Results} 
\label{res}
We have computed the detection efficiency for our toy setup through the three independent simulation chains presented in section \ref{principle}. Detection efficiency is defined here as the ratio of the number of showers detected to the total of $100$ selected tau decays. The parameters ranges explored initially are distances $D=\{20, 30, 40, 60, 80, 100\}$ km and slopes  $\alpha=\{0, 5, 10, 15, 20, 45, 90\}$ degrees. This coarse step is mainly motivated by computation time and disk space considerations for the {\it microscopic} simulation. 

We first show a relative comparison of the different methods before discussing the effects of the topography on the detection efficiency.
\begin{figure*}[tb]
\center
\includegraphics[width=0.49\textwidth]{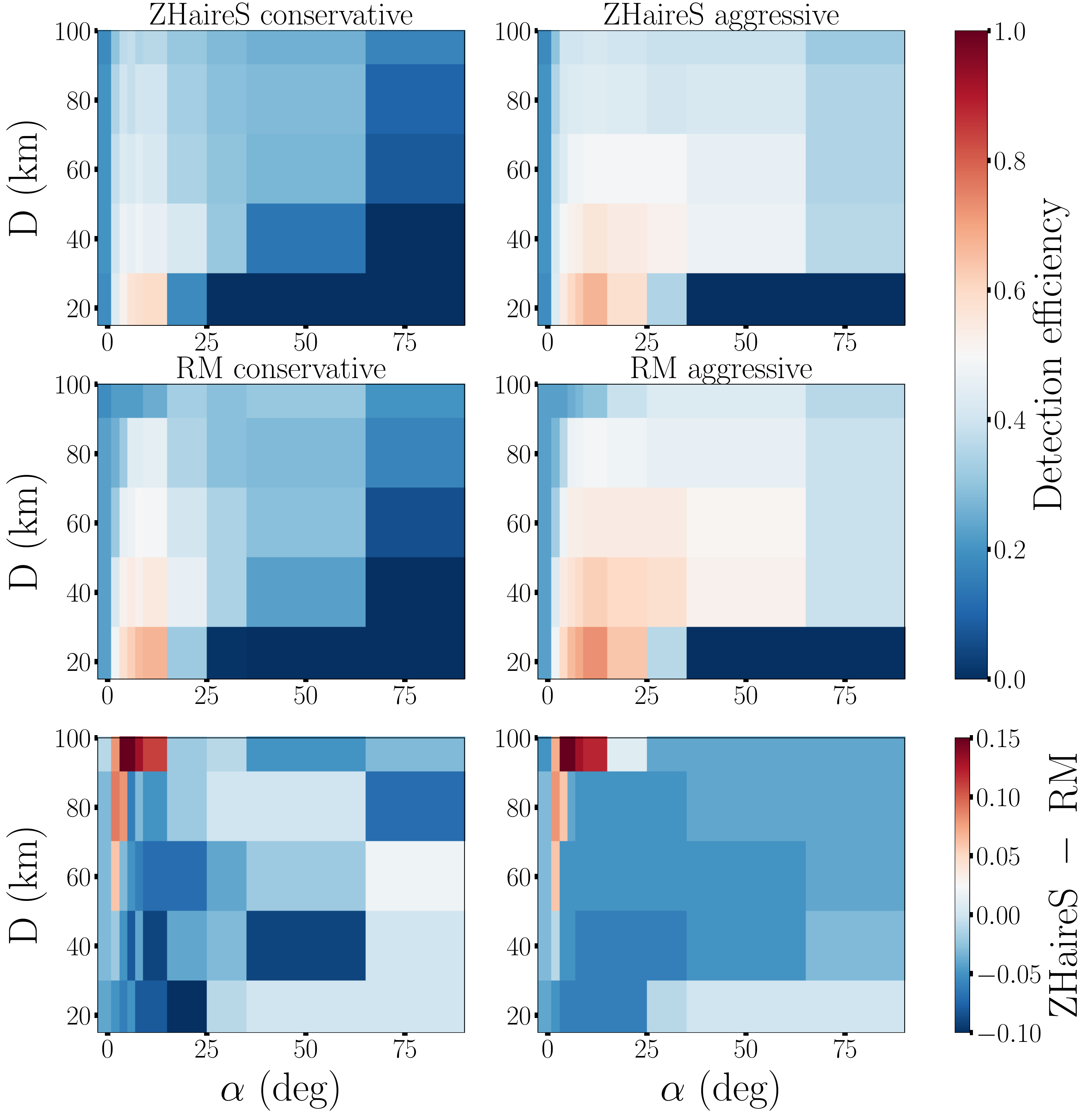}
\includegraphics[width=0.49\textwidth]{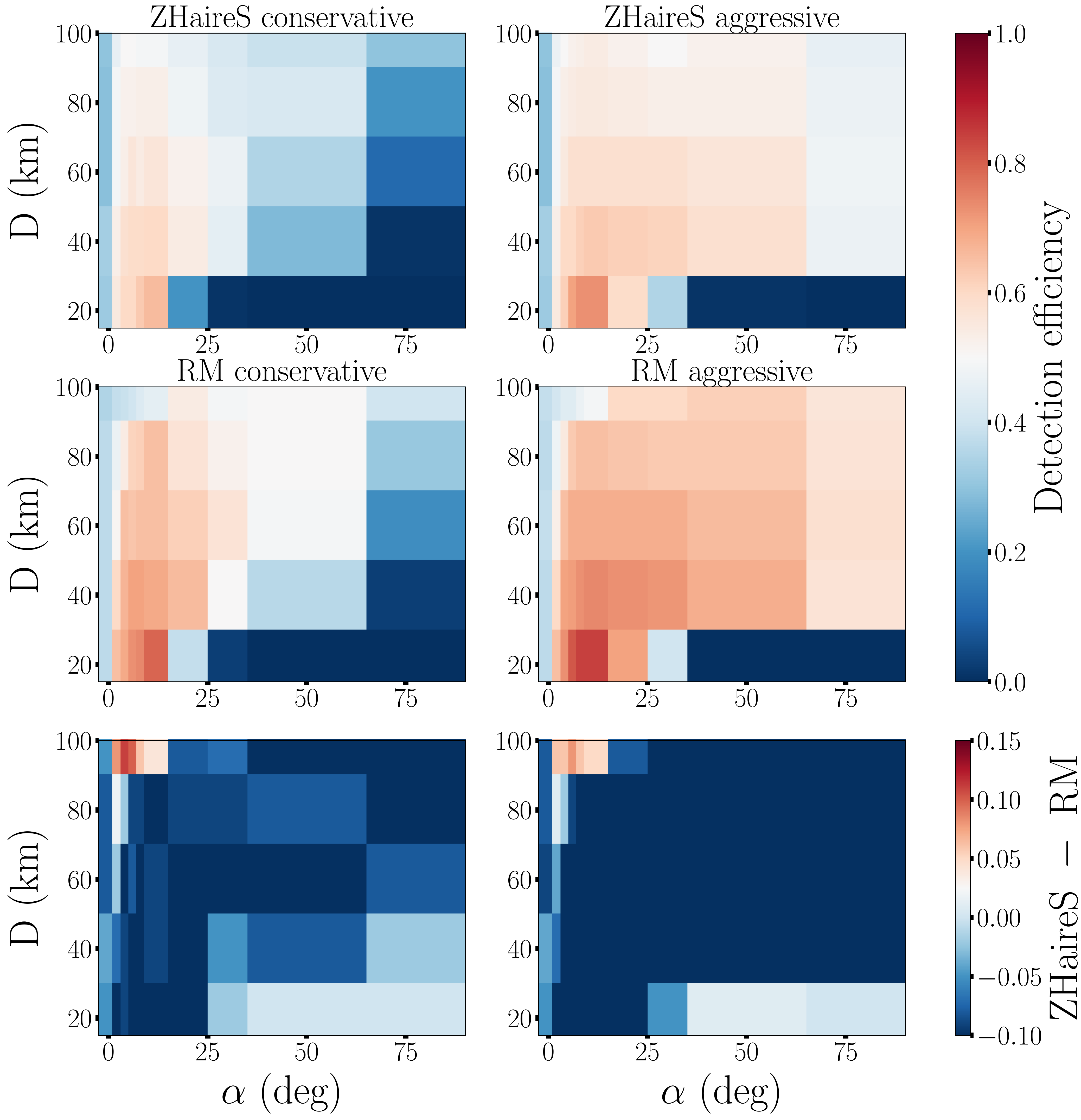}
\caption{{\it Left:} Detection efficiency as a function of the distance $D$ and slope $\alpha$ for the simulation set with a primary neutrino energy of $10^9$\,GeV. Comparison between ZHAireS and {\it Radio Morphing} (respectively {\it top} and {\it middle} plots, while the difference is plotted at the {\it bottom}) and conservative thresholds ({\it left}) and aggressive thresholds ({\it right}). {\it Right:} Same for a primary neutrino energy of $10^{10}$\,GeV.}
\label{Zhaires_RM_sens}
\end{figure*}
%

\begin{figure*}[tb]
\center
\includegraphics[width=0.49\textwidth]{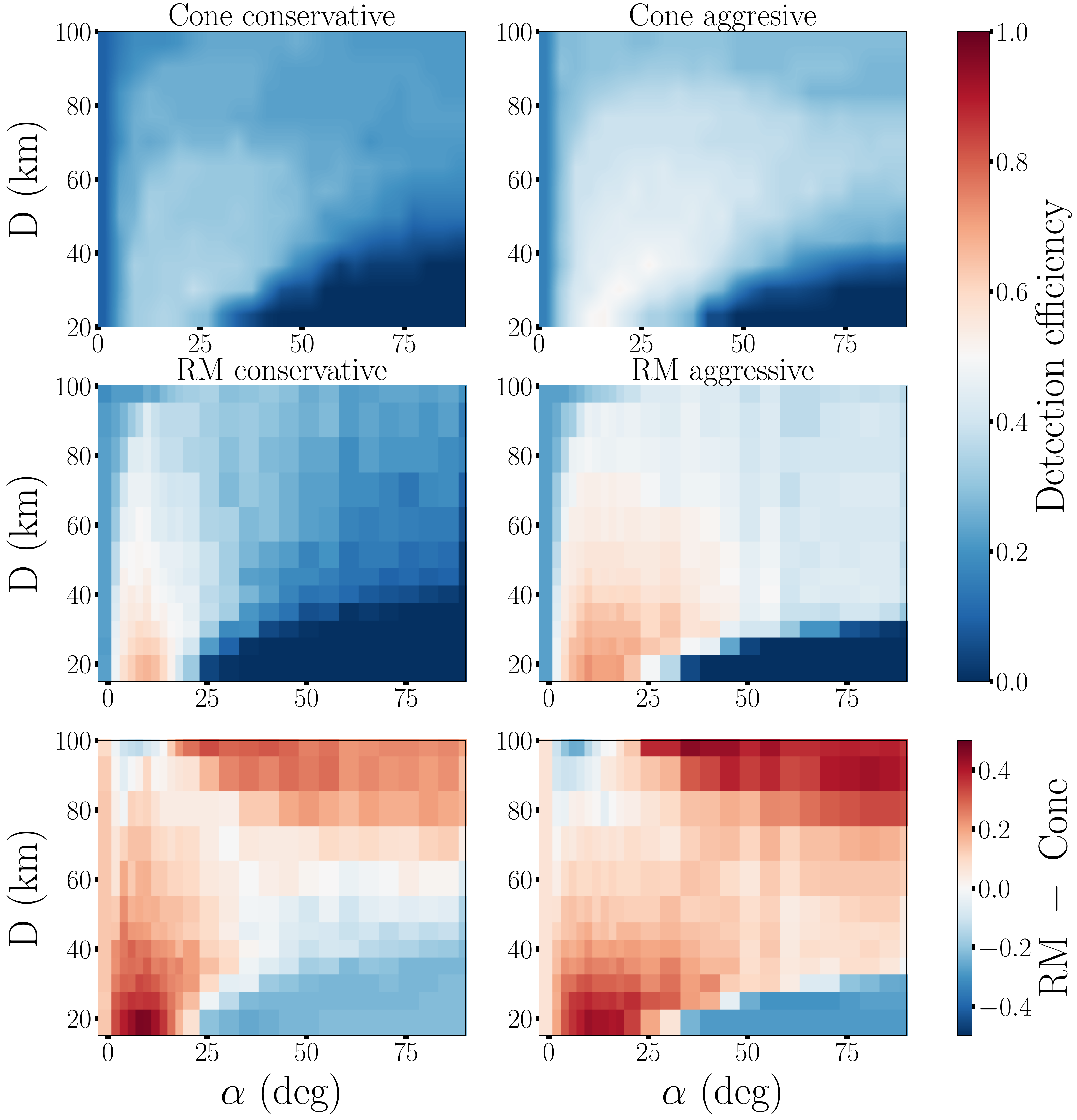}
\includegraphics[width=0.49\textwidth]{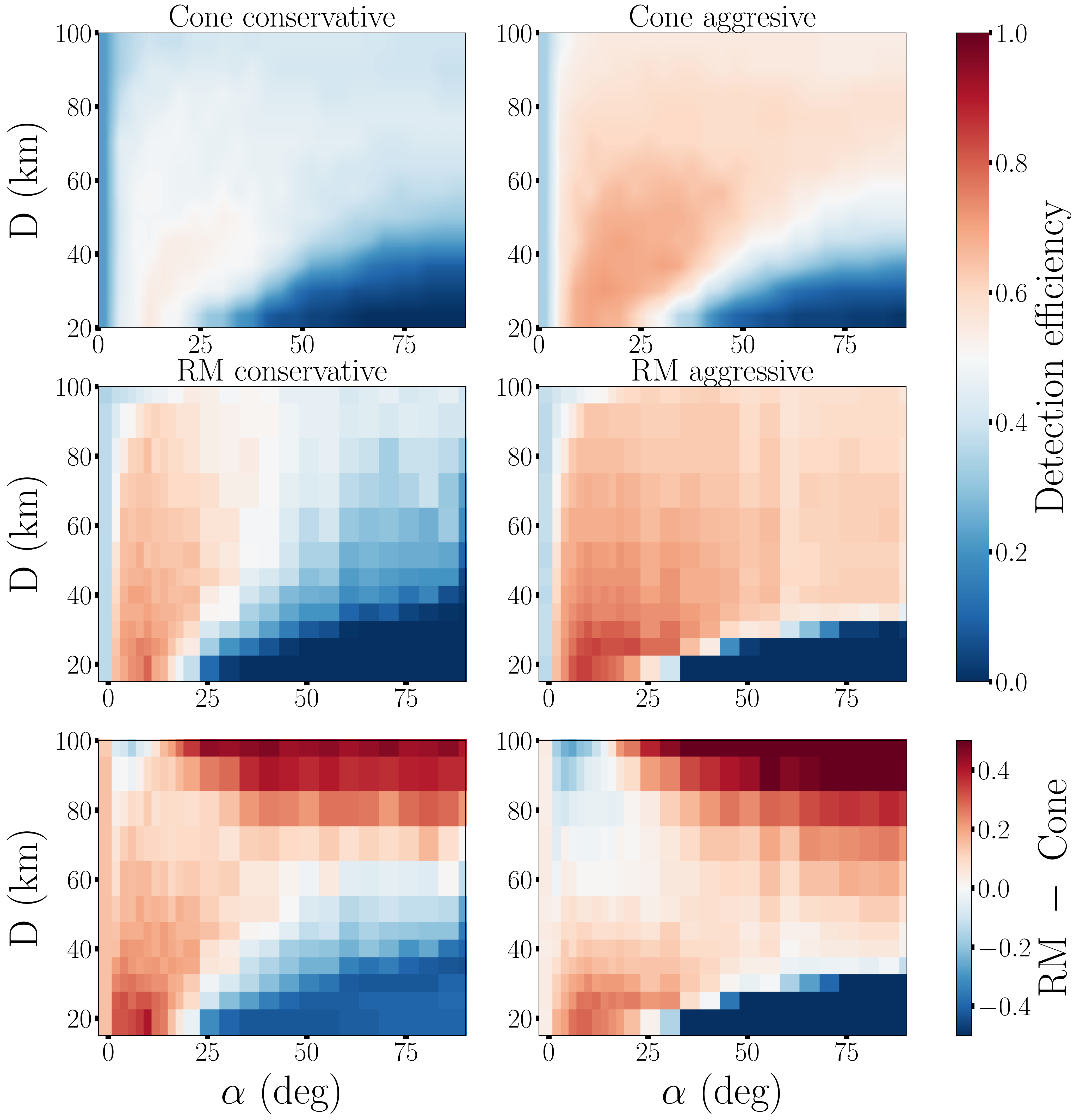}
\caption{ {\it Left:} Detection efficiency as a function of distance $D$ and slope $\alpha$ for the simulation set with a primary neutrino energy of $10^9$\,GeV. Results are plotted for the {\it Cone Model} ({\it top}) and {\it Radio Morphing} ({\it middle}), as well as the difference ({\it Radio-Morphing} - {\it Cone Model}) ({\it bottom}). Conservative ({\it left}) and aggressive ({\it right}) threshold hypothesis are also considered ({\it right}). {\it Right:} Same for a primary neutrino energy of $10^{10}$\,GeV.}
\label{Cone_RM_sens}
\end{figure*}

\begin{figure*}[tb]
\center
\includegraphics[width=0.49\textwidth]{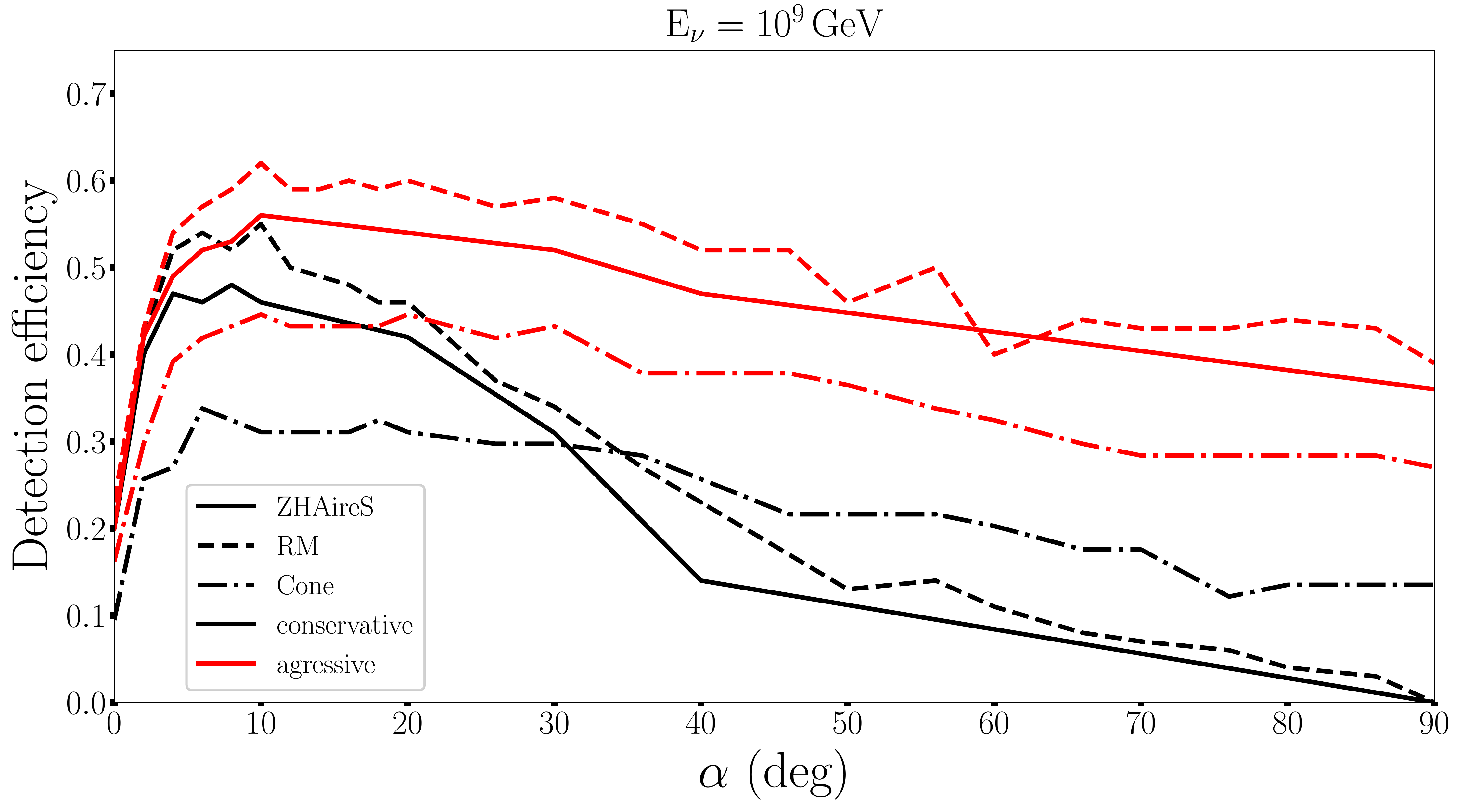}
\includegraphics[width=0.49\textwidth]{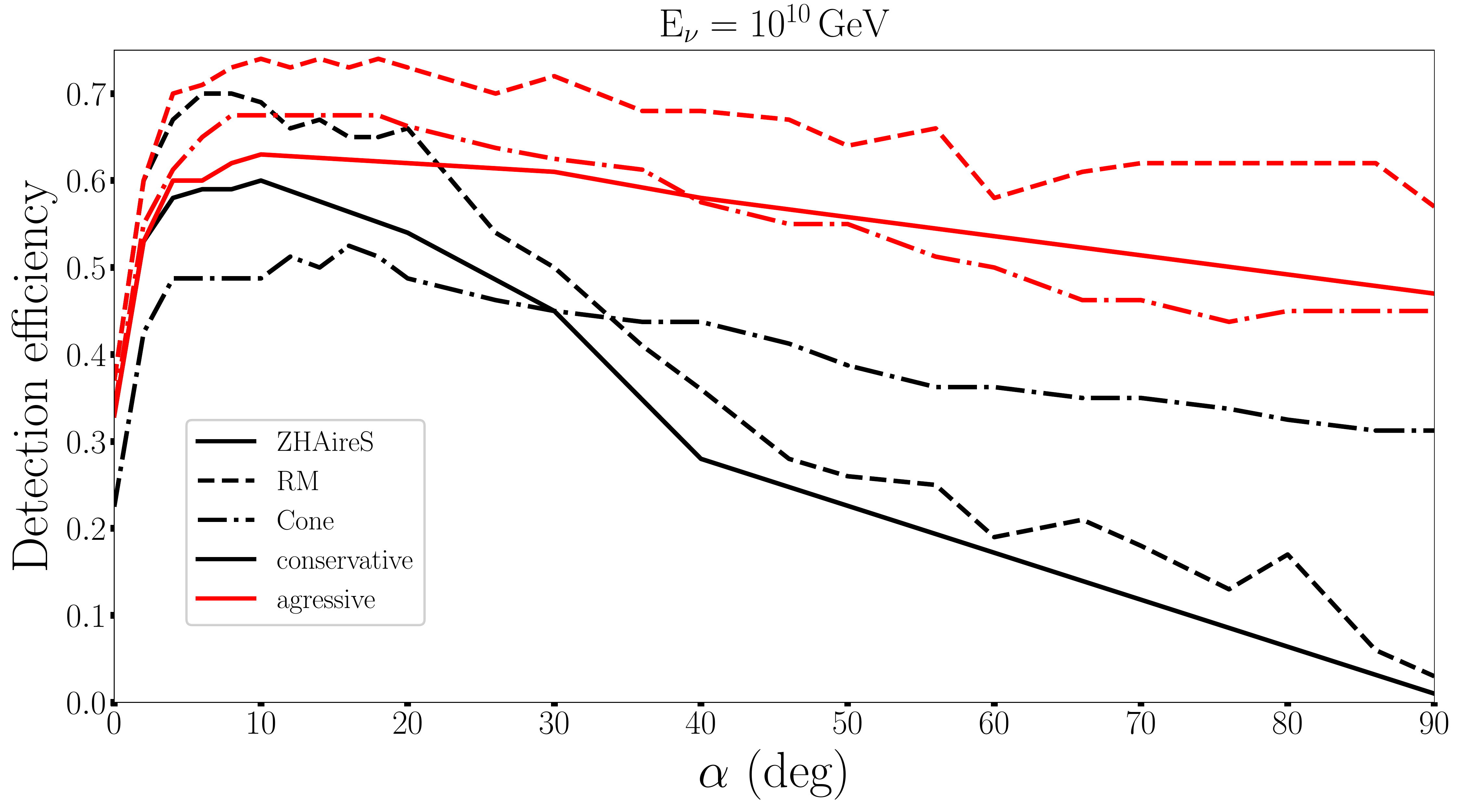}
\includegraphics[width=0.49\textwidth]{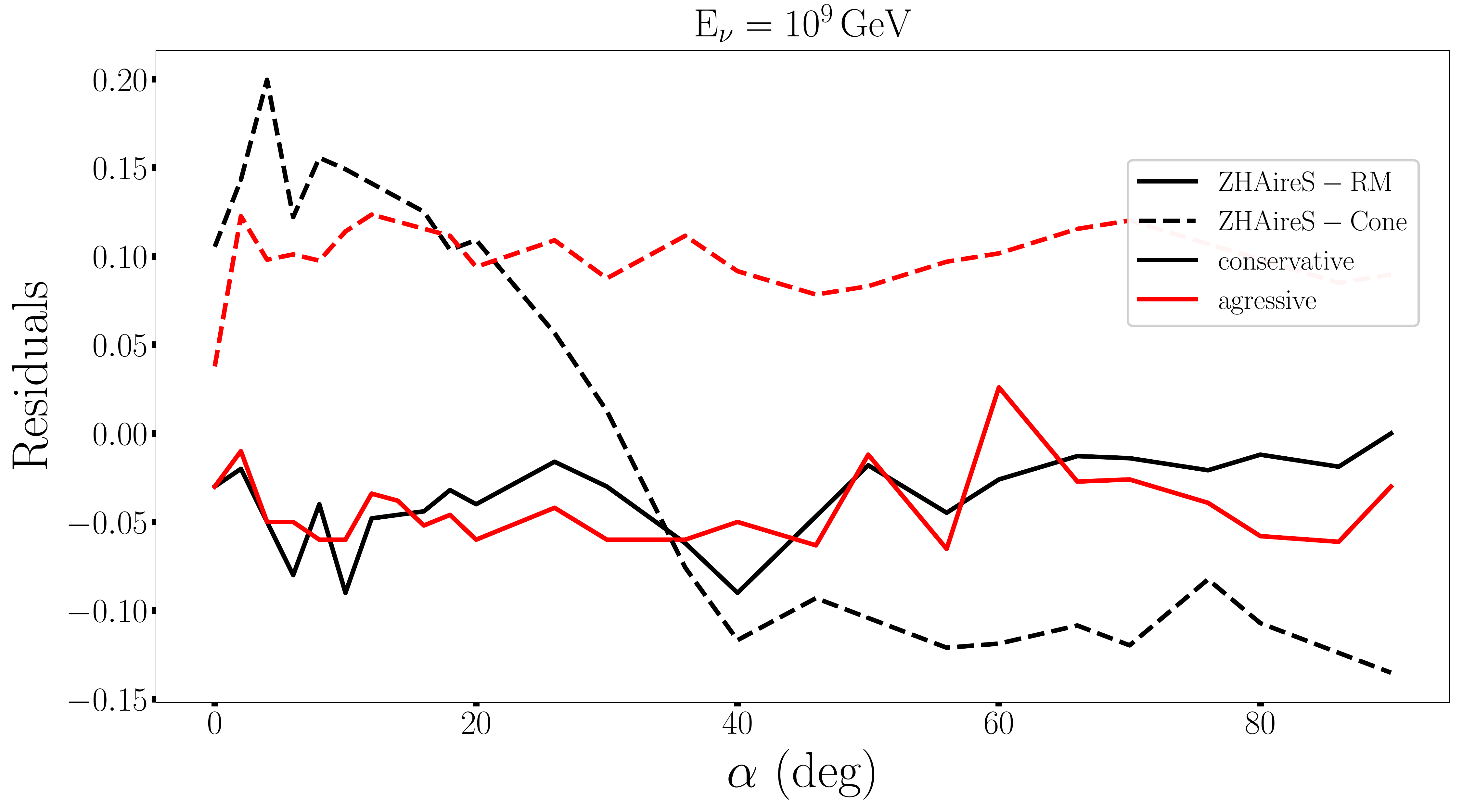}
\includegraphics[width=0.49\textwidth]{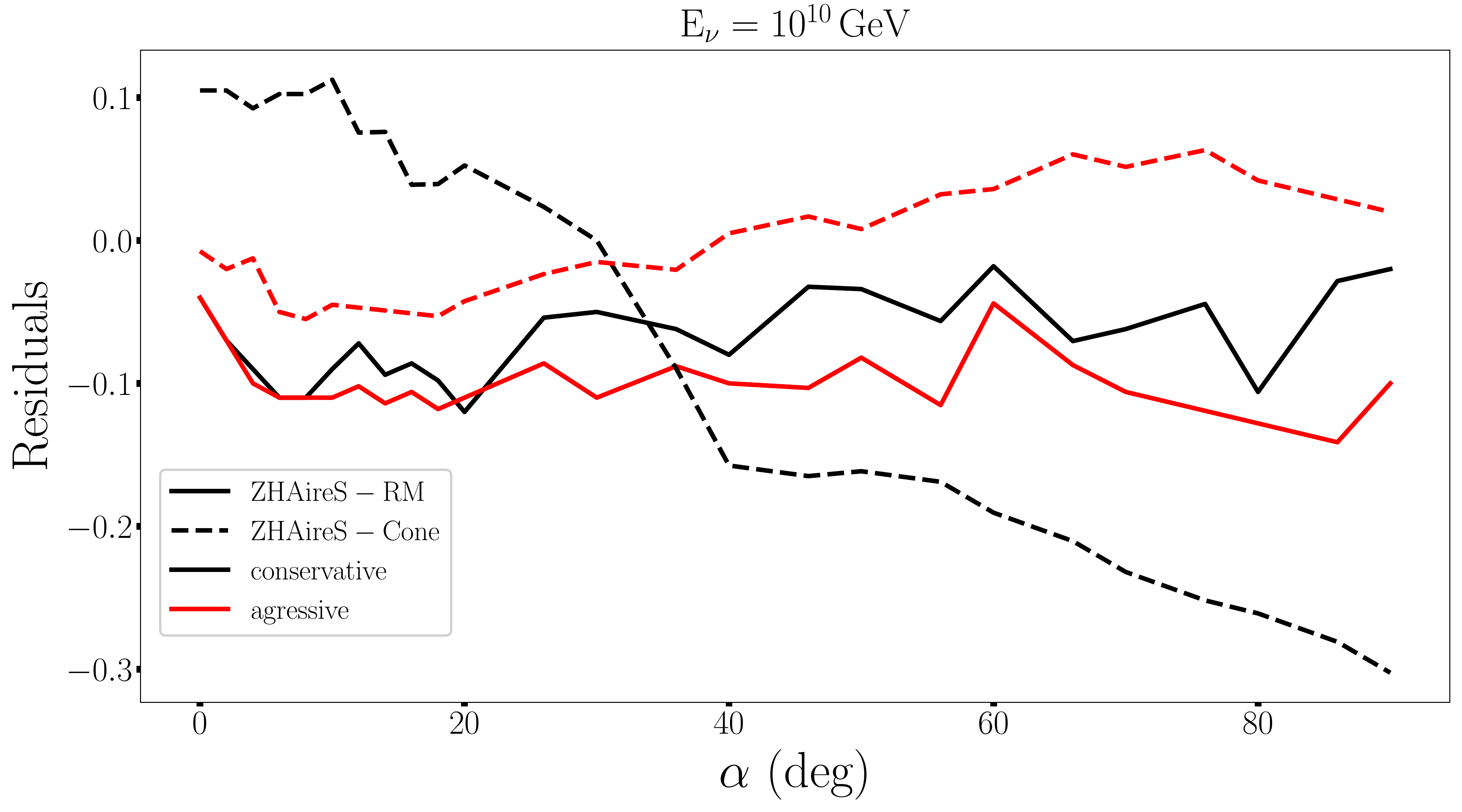}
\caption{{\it Top:} Detection efficiency as a function of slope $\alpha$ for a distance $D= 40$\,km for neutrinos energies of $10^9$ ({\it left}) and $10^10$\,GeV ({\it right}). Comparisons between the microscopic (solid lines), {\it Radio Morphing} (dashed lines) and {\it Cone Model} (dash-dotted lines), for conservative (black lines) and aggressive (red lines) threshold hypothesis. {\it Bottom:} Differences {\it ZHAireS-RadioMorphing} and {\it ZHAireS - Cone Model} for the data shown in the top pannel, following the same color code.} 
\label{distance_slices_1e18}
\end{figure*}

\subsection{Relative comparison}

Figure \ref{Zhaires_RM_sens} shows that the  {\it Radio Morphing} treatment induces trigger efficiencies at most 15\% higher than {\it microscopic} simulations.
This confirms results obtained in \cite{Zilles:2018kwq} and qualifies the {\it Radio Morphing} chain as a valid tool for the study presented in this article. Taking advantage of the factor $\sim$100 gain in computation time of {\it Radio Morphing} compared to {\it microscopic} simulations\,\cite{Zilles:2018kwq} we then decrease the simulation step size down to 2$^{\circ}$ for slope $\alpha$ and 5\,km for the distance to decay $D$, allowing for a more detailed study of the effect of topography on the array detection efficiency.

This refined analysis is presented in Figure \ref{Cone_RM_sens}, where results of the {\it Cone Model} are also shown. The distribution of the {\it Cone Model} detection efficiency in the ($\alpha$,$D$) plane follows a trend similar to the {\it Radio Morphing} one, with differences within $\pm$20\% for most of the parameter space ($\alpha,D$). There are however some differences, in particular a significant under-estimation with the {\it Cone Model} in the ranges ($\alpha>30\degree$, $D>80$\,km) and ($\alpha<20\degree$, $D<30$\,km). There is also a flatter distribution as a function of slope for the conservative trigger hypothesis, which results in an over-estimation for ($\alpha>30\degree$, $D<40$\,km) for the {\it Cone Model}, also visible on Figure \ref{distance_slices_1e18}. 

Discrepancies are not surprising since the {\it Cone Model} is an approximate method as already pointed out in section \ref{cone}.  One should however be reminded that slopes $\alpha>30\degree$ correspond to extreme cases, very rare in reality and which cannot be considered for actual deployment. For realistic slope values $\alpha<30\degree$, the {\it Cone Model} detection efficiencies differ from those of the {\it microscopic} approach by -30\% at most.  The {\it Cone Model} can thus safely be used to provide in a very short amount of time a rough and conservative estimate of the neutrino sensitivity for realistic topographies. This result also provides an {\it a posteriori} validation of the initial computation of the GRAND array sensitivity \cite{Martineau:2016yj}, even though the cone was then parametrized from showers simulated in the 30-80\,MHz frequency range. 

\subsection{Toy-setup discussion}
\label{toymodelres}
Below we study how the topography affects the detection potential of neutrino-induced air showers by a radio array. To do that, we use the results of the {\it Radio Morphing} chain, which provide at the same time good reliability and fine topography granularity as explained in the previous section.

Despite statistical fluctuations obviously visible in Figures \ref{Cone_RM_sens} and \ref{distance_slices_1e18}, general trends clearly appear. Four striking features can in particular be singled-out: \\
$\bullet$ A significant increase of the detection efficiency for slopes varying from 0 degree up to few degrees: the detection efficiency for a flat area is  lower by a factor $3$ compared to an optimal configuration $\qty(\alpha, D) \approx \qty(10 \degree, 25\,{\rm km})$. This result is consistent with the study presented in \cite{GRANDWP}, where the effective area computed for a real topography on a mountainous site was found to be four times larger than for a flat site. \\
$\bullet$ Limited variation of the detection efficiency for slopes between $\sim2\degree$ and $\sim20\degree$. \\
$\bullet$ An efficiency slowly decreasing for slopes larger than $\sim20\degree$. This is in particular valid for distances $D$ shorter than 40\,km, where the detection efficiency is nearly null. \\
$\bullet$ A slow decrease of the detection efficiency with increasing value of $D$. 

\begin{figure*}[tb]
\center
\includegraphics[width=0.49\textwidth]{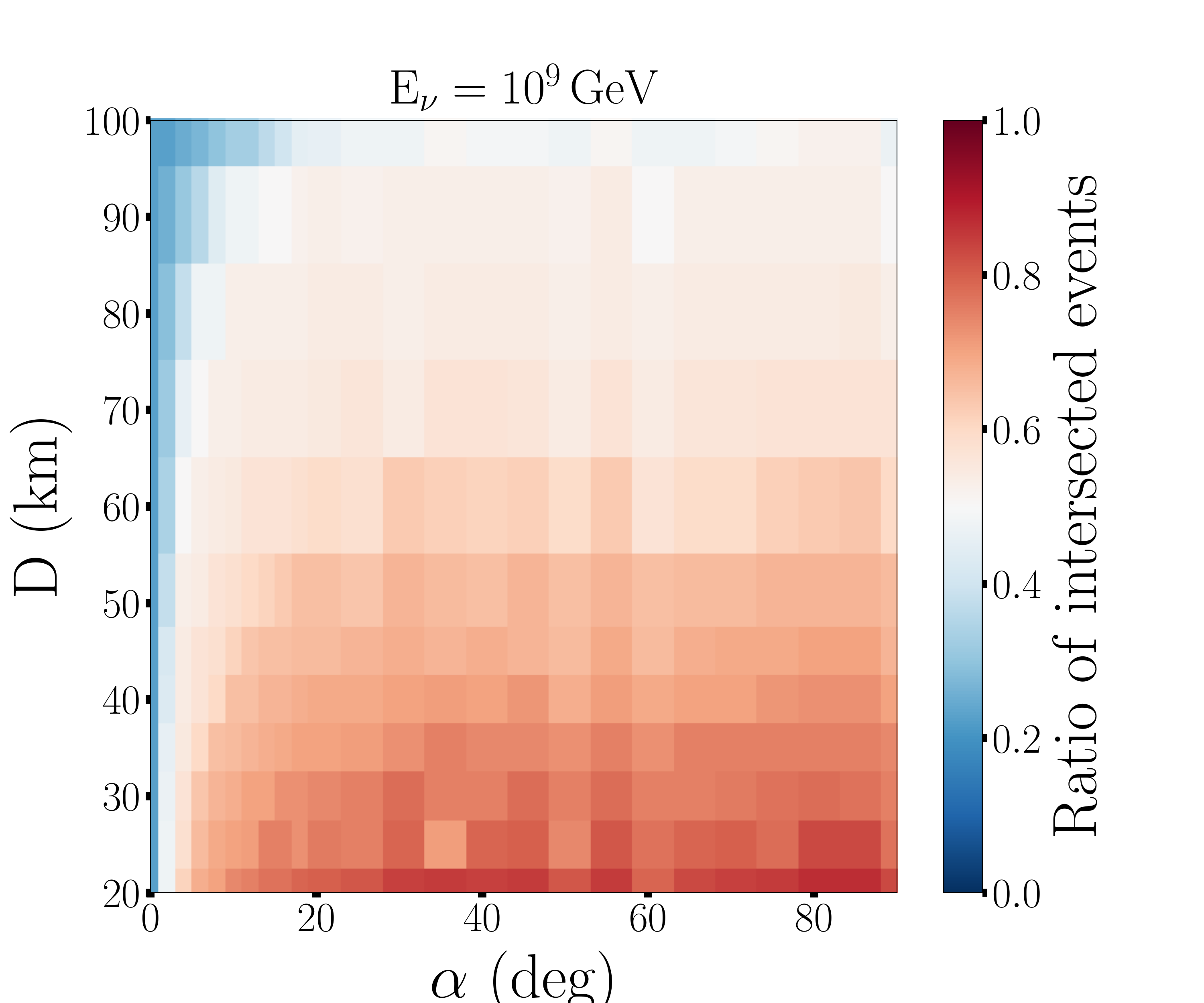}
\includegraphics[width=0.49\textwidth]{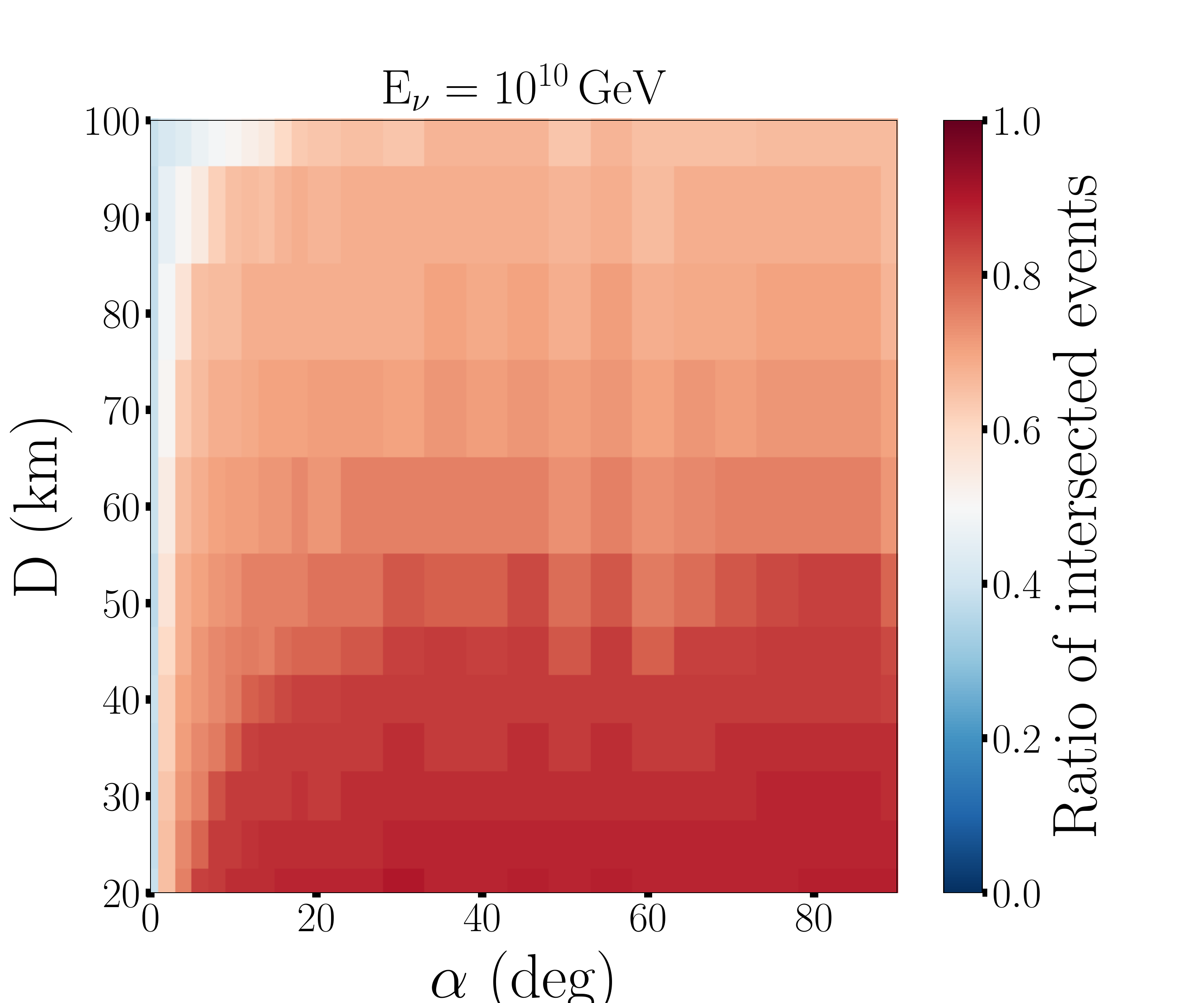}
\caption{ {\it Left:} Fraction of events intersecting the detection area as a function of distance $D$ and slope $\alpha$ for the simulation set with a primary neutrino energy of $10^9$\,GeV. {\it Right:} Same for a primary neutrino energy of $10^{10}$\,GeV.} 
\label{fly_above}
\end{figure*}

\medskip
To interpret these results, we may first consider that two conditions have to be fulfilled to perform radio-detection of showers: first the radio beam must hit the detector, then enough antennas (five in this study) have to trigger on the corresponding radio signal. In order to disentangle these two factors ---one mostly geometrical, the other experimental---, we display in Figure \ref{fly_above} the fraction of events reaching the detector as a function of the parameters $\qty(\alpha, D)$. These events are defined by a non-null intersection between the detector plane and a $3\degree$ half-aperture cone centered on the shower trajectory, a conservative and model-independent criterion. 

It appears from Figure \ref{fly_above} that the large fraction ---around 90\%--- of showers flying above the detector is the main cause of the limited efficiency of a flat detection area. As a corollary, the steep rise of detection efficiency with increasing slope is clearly due to the increasing fraction of intercepted showers. Figures \ref{Cone_RM_sens} and \ref{fly_above} however differ significantly for configurations corresponding to $\alpha>20\degree$ and $D<40$\,km: the fraction of intercepted events varies marginally with $\alpha$ at a given $D$, while the detection efficiency drops. This means that the first condition for detection ---detector inside the radio beam--- is fulfilled for these configurations, but the second ---sufficient number of triggered antennas--- is not, because the tau decay is too close, and the radio footprint at ground consequently too small. 
The situation may be compared ---with a 90$\degree$ rotation of the geometry of the problem--- to the radio-detection of "standard" air showers with zenith angle $\theta<60\degree$, which suffers from limited efficiency for sparse array\,\cite{Charrier:2018fle}. A larger density of detection units would certainly improve detection efficiency, but the need for large detection areas, imposed by the very low rate of neutrino events, discards this option. 

Finally the slow decrease in efficiency with increasing value of $D$ is mostly due to geometry, as the fraction of intersecting events diminishes with $D$ in similar proportion.

\begin{figure*}[tb]
\center
\includegraphics[width=0.49\textwidth]{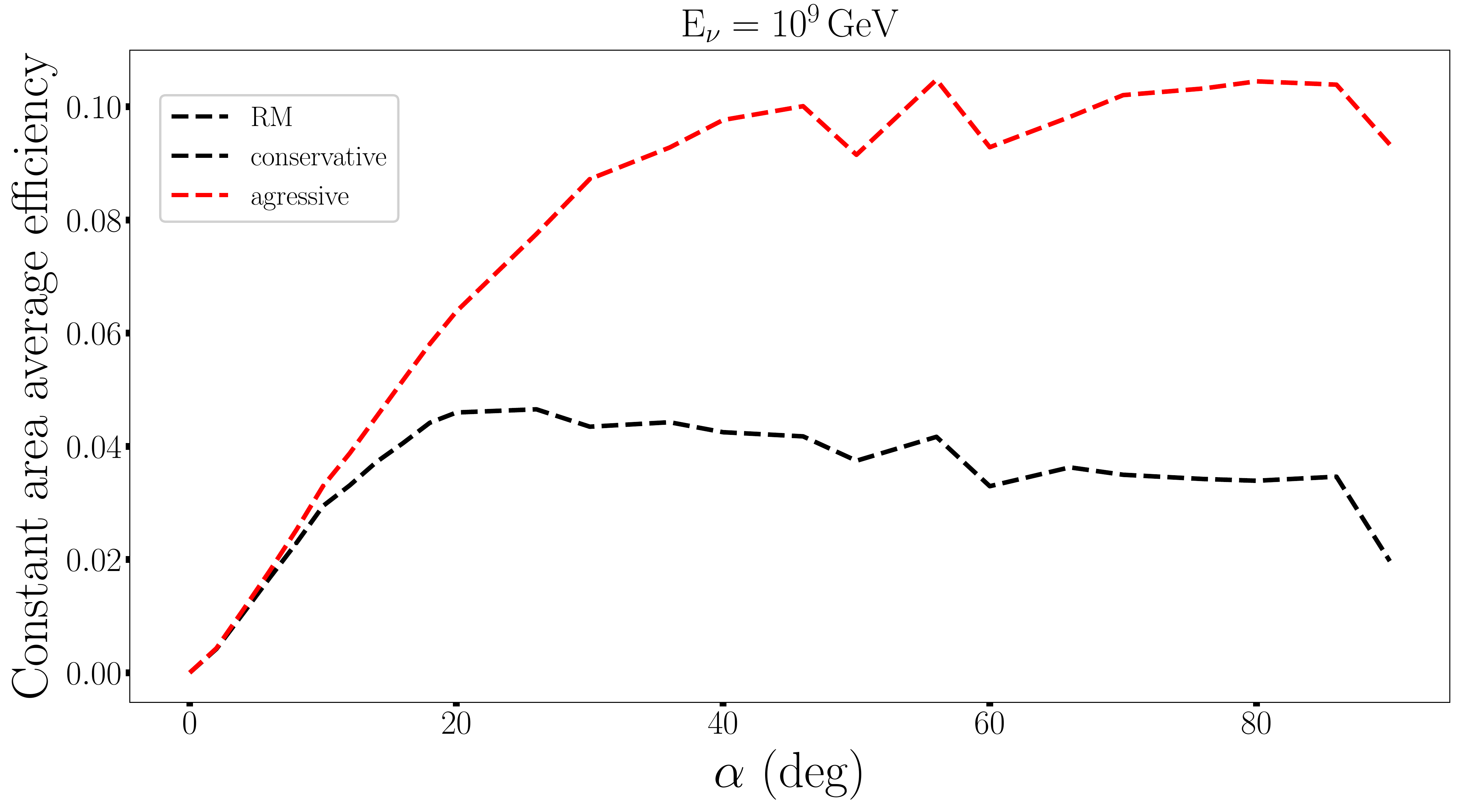}
\includegraphics[width=0.49\textwidth]{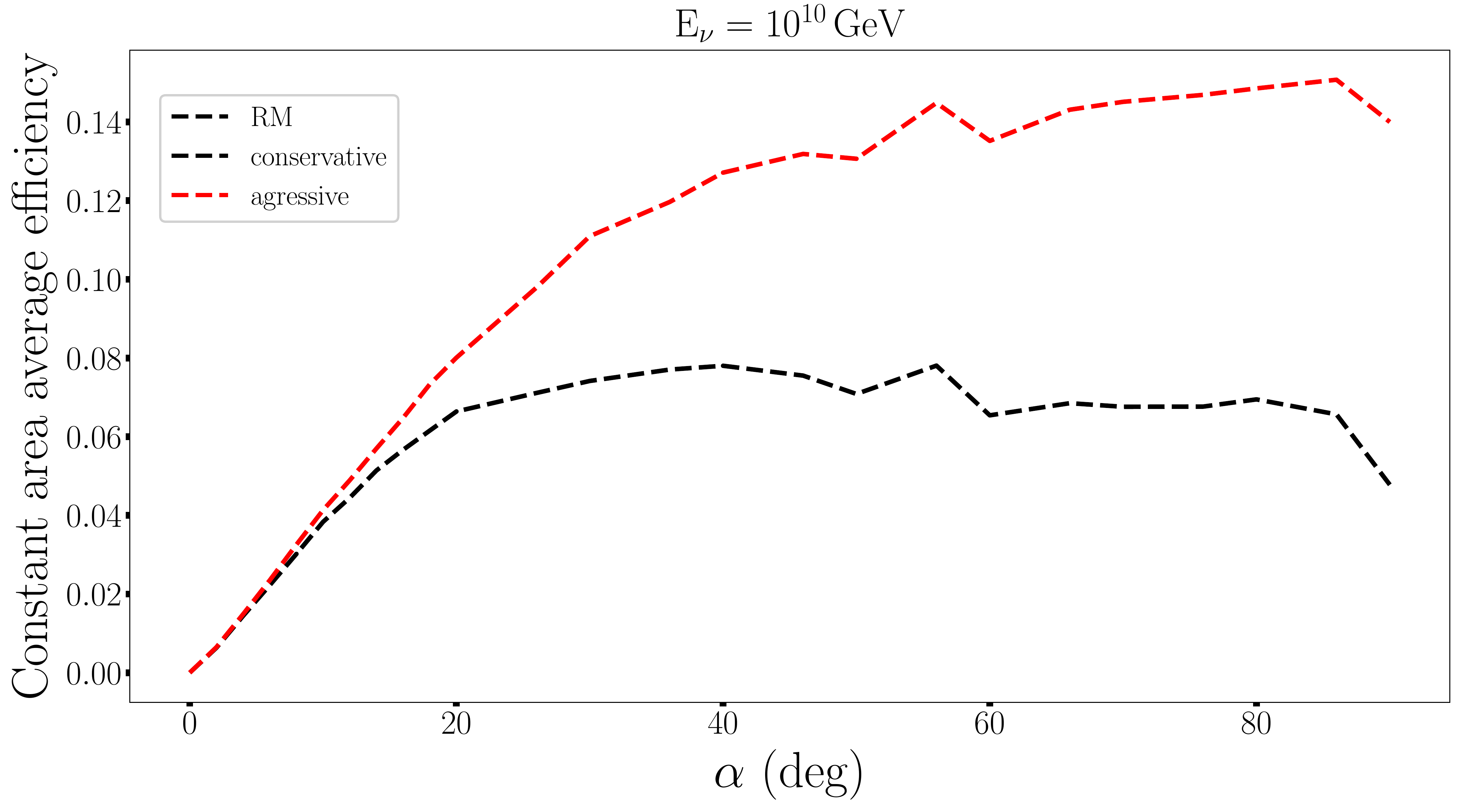}
\caption{{\it Left:} Average detection efficiency over a constant detector area as a function of slope for the simulation set with a primary neutrino energy of $10^9$\,GeV. {\it Right:} Same for a primary neutrino energy of $10^{10}$\,GeV. In both cases values are computed with the {\it Radio Morphing} treatment.}
\label{constant_average_detection_eff}
\end{figure*}
\medskip

Yet, one could argue that this result is biased by the detector layout defined in our toy-setup. The infinite width of the detection plane combined with a limit on the detector elevation ($3000$\,m above the reference altitude, see section \ref{toymodeldescription})  indeed implies that a detector deployed over mild slopes is larger than one deployed over steeper ones in this study. A value $\alpha =10\degree$ for example allows for a detector extension of $3/\sin\alpha\sim17$\,km, while $\alpha = 70\degree$ implies a value six times smaller. Considering a constant detector area for all configurations ($\alpha$,D) and comparing their effective area ---or expected event rates--- would avoid such bias, but would require a complete Monte-Carlo simulation. This is beyond the scope of this study, and would be useful only if real topographies were taken into account.

It is however possible to estimate this bias by studying how the {\it constant area detector efficiency} varies with slope. This quantity is defined as the detection efficiency averaged over $D$ and weighted with a factor $\sin\alpha$. As $D$ measures the amount of empty space in front of the detector (see section \ref{toymodeldescription}), averaging the efficiency over all values of $D$ allows us to take into account all possible shower trajectories for a given slope value. The factor $\sin\alpha$ corrects for the variation of the detector area with slope. The {\it constant area detector efficiency} can therefore be understood as a proxy for the event rate per unit area of a detector deployed on a plane of slope $\alpha$, facing an infinite flat area. The {\it constant area average efficiency} computed from the {\it Radio Morphing} results is displayed as a function of slope on Figure \ref{constant_average_detection_eff}. \\
Beyond a certain threshold ($\sim20\degree$ for the conservative case, $\sim30\degree$ for the aggressive one), there is no significant variation of its value with $\alpha$, because the poor performance of steep slopes for close-by showers (i.e. small values of $D$) compensates for the larger area factor $\sin\alpha$. Figure \ref{constant_average_detection_eff} also confirms the clear gain of a slope ---even mild--- compared to a detector deployed over flat ground. \\
Only showers propagating towards the slope were considered in this study, but one can deduce from Figures \ref{distance_slices_1e18} and \ref{fly_above} that the opposite trajectory (corresponding to a down-going slope, i.e. $\alpha<0$) results in a $\sim0$ detection probability. For showers traveling transverse to the slope inclination (i.e. along the East-West axis in our configuration), basic geometric considerations allow to infer that the situation is probably comparable to horizontal ground. The boost effect of value 3 determined for showers propagating towards the detector plane thus certainly corresponds to a best-case scenario. Computing the net effect of a non-flat topography on the neutrino detection efficiency for random direction of arrivals cannot be performed with this toy-setup configuration (see section \ref{toymodeldescription} for details). However we note that a study presented in Ref. \cite{GRANDWP} points towards a boost factor of $\sim$2 on the detection of upward-going showers for the specific site used in that work.

\section{Conclusion}
We have studied the impact of the topography for radio-detection of neutrino-induced Earth-skimming air showers. For this purpose, we have developed a toy setup with a simplified topography for the detector, depending on two parameters: the distance between the air shower injection point and the detector array, and the ground slope of the detector array. We have computed the neutrino detection efficiency of this toy detector configuration through three computation chains: a microscopic simulation of the shower development and its associated radio emission, a radio-signal computation using {\it Radio Morphing} and an analytical treatment based on a {\it cone model} of the trigger volume. 

The comparison of these three independent tools confirms that {\it Radio Morphing} is a reliable method in this framework, while the {\it Cone Model} offers a fast, conservative estimate of the detection efficiency for realistic topographies. 
The latter can thus be used to perform in a negligible amount of time a  preliminary estimate of the potential for neutrino detection of a given zone, and the former can then be used to carry out a detailed evaluation of selected sites instead of full Monte-Carlo simulations.

More importantly, the results presented here show that ground topography has a great impact on the detection efficiency, with an increase by a factor $\sim$3  for angles of just a few degrees compared to a flat array and for an optimal case where shower trajectories face the detector plane. This boost effect is very similar for any slope value ranging between $2\degree$ and $20\degree$. The other noticeable result of this study is the moderate effect of the distance on the detection efficiency, with comparable values for tau decays taking place between $20$ and $100$\,km from the detector.

Two slopes facing each other with tens of kilometers between them may therefore  constitute the optimal configuration for neutrino detection, as they would correspond to enhanced rates for the two directions perpendicular to the slopes. Wide valleys or large basins could offer such topographies
and will consequently be primarily targeted in the search for the optimal sites where the $\mathcal{O}$(10) sub-arrays composing the GRAND array could be deployed to optimize its neutrino detection efficiency. An effort in this direction has been initiated in the framework of the GRAND project.

\subsection*{Acknowledgments}
We are grateful to Clementina Medina and Jean-Christophe Hamilton for suggesting this study. This work is supported by the APACHE grant (ANR-16-CE31-0001) of the French Agence Nationale de la Recherche. We also thank the France China Particle Physics Laboratory for its support. The simulations were performed using the computing resources at the CC-IN2P3 Computing Centre (Lyon/Villeurbanne – France),  partnership between CNRS/IN2P3 and CEA/DSM/Irfu.
\section{References}
\bibliographystyle{elsarticle-num} 
\bibliography{biblio} 

\end{document}